\documentclass[preprint,pre]{revtex4-2}
\usepackage{graphicx}
\usepackage{lipsum}
\usepackage{float}
\usepackage{amsmath}
\usepackage{amssymb}
\usepackage{mathtools}
\usepackage{mathrsfs}

\begin{document}

\title{First-passage processes in a deterministic one-dimensional  \\
cellular automaton model of traffic flow}

\author{Ofer Biham, Gilad Hertzberg Rabinovich and Eytan Katzav} 
\affiliation{Racah Institute of Physics, The Hebrew University, Jerusalem 9190401, Israel}

\begin{abstract}

We present analytical results for first-passage processes
in a deterministic one-dimensional (1D) cellular automaton (CA) model of traffic flow.
Starting at time $t=0$ from a random initial state with car density $p$, 
at every time step $t \ge 1$ each car moves one step to the right if the cell 
on its right is empty,
and is stopped if it is occupied by another car.
The model, which coincides with CA rule 184 in Wolfram's numbering scheme,
exhibits a continuous dynamical phase transition at $p=1/2$,
between a low-density free-flowing phase and a high-density
congested phase.
Using the framework of first-passage processes,
we derive a closed-form expression for the distribution 
$P(T_{\rm FS}=t)$
of first-stopping (FS) times,
which is the probability that a randomly selected car will be stopped for the 
first time at time $t$.
We also obtain a closed-form expression for the 
stopping probability $P_{\rm S}(t)$, which is the probability that a randomly selected car
will be stopped at time $t$.
In the low-density phase of $0 < p < 1/2$, 
the probability $P_{\rm S}(t)$ yields a closed-form expression for  
the distribution 
$P(T_{\rm LS}=t)$
of last-stopping (LS) times,
which is the probability that a randomly selected car will be stopped for the
last time at time $t$, beyond which it will move freely indefinitely.
In this regime, we analyze the relation between the LS time and the 
number of stopping events $N_{\rm S}$ 
which take place up to that time.
We present closed-form expressions 
for the joint distribution $P(T_{\rm LS}=t,N_{\rm S}=n)$,
for the two conditional distributions that emanate from it
and for the marginal distribution $P(N_{\rm S}=n)$.
These results provide insight on the time scales of congestion
and relaxation in deterministic traffic flow
from the point of view of individual cars.
In a broader context, they provide insight on complex relaxation processes
that involve many interacting particles, 
such as the processes that appear in deterministic surface growth.

\end{abstract}


\maketitle
\newpage

\section{Introduction}

Cellular automaton (CA) models 
\cite{Wolfram1983,Wolfram1984}
provide a powerful framework for the analysis of traffic flow processes
\cite{Biham1992,Maerivoet2005}.
These models simulate the movement of vehicles on a discretized space,
where each cell can be occupied by at most one vehicle.
They evolve in discrete time steps, with vehicles following a set of 
deterministic or probabilistic rules for acceleration, deceleration and lane changing.
The models capture essential traffic phenomena such as stop-and-go waves, 
the buildup of congestion and phase transitions between free-flowing traffic and jammed traffic 
\cite{Chowdhury2000,Helbing2001,Schadschneider2002,Nagatani2002}.

The Nagel-Schreckenberg 
model
\cite{Nagel1992}
is a stochastic CA model
that captures essential features of highway traffic on a single lane,
such as the spontaneous formation of traffic jams and a phase transition between free and congested flow 
\cite{Nagel1993,Schadschneider1993,Schreckenberg1995}.
In this model the velocities take integer values and the dynamics includes acceleration, slowing-down
and randomization.
In particular, the randomization step is crucial in order to capture realistic traffic behavior
and to account for the irregularity and unpredictability of human behavior.
The Nagel-Schreckenberg
model established a minimal microscopic framework for jam formation 
in highway traffic. 
This framework was later extended by introducing 
speed synchronization,  
which led to a generalized phase structure comprising 
free flow, synchronized flow, and wide moving jams, in agreement with 
empirical results 
\cite{Kerner1996,Kerner2002}.

Taking the deterministic limit of the Nagel-Schreckenberg
model and restricting the velocities to $0$ and $1$
yields a simple one-dimensional (1D) model, referred to below as the 
deterministic simple traffic-flow (DSTF) model.
It starts at time $t=0$
from a random initial state,
in which each cell is occupied
by a car with probability $p$ or is empty with probability $1-p$.
Every time step $t \ge 1$,
each car moves one step to the right if the cell on its right is empty,
and is stopped if it is occupied by another car.
These dynamical rules, 
which coincide with CA rule 184
in Wolfram's numbering scheme
\cite{Wolfram1983,Wolfram1984},
were studied in the contexts of 
surface growth
\cite{Krug1988},
traffic flow 
\cite{Sasvari1997} 
and chemical reactions 
\cite{Belitsky2005}.
An important property of 
the model,
which was proved in the context of 
CA rule 184,
is that the number of cars is conserved
\cite{Boccara1998,Boccara2002}.
The pattern formation in CA rule 184 
as time evolves was studied in Refs.
\cite{Fuks1997,Fuks1999,Fuks2023}.

The DSTF model exhibits a 
continuous dynamical
phase transition at $p=1/2$, separating between  
the low-density phase of $0 < p < 1/2$ 
and the high-density phase of $1/2 < p < 1$.
In the low-density phase, 
on a finite lattice with periodic boundary conditions,
the system evolves toward 
a free-flowing periodic (FFP) state.
This is a stationary state in which
every pair of consecutive cars is separated by at least one empty cell
and the velocity of all the cars is $v=1$.
In the high-density phase the system evolves toward a steady-state
in which jammed and free-flowing traffic coexist 
and the average velocity is $\langle v \rangle = (1-p)/p$.
The phase transition at $p=1/2$ was rigorously characterized using phase-separating
invariant measures for the low and high-density phases
\cite{Belitsky2005}.
Combining the results for the low-density and high-density phases,
it is found that under steady-state conditions
the flow, which is the product of the
car density and the average velocity, is given by
$J(p) = \min(p,1-p)$.
Until recently, most studies of CA rule 184 and its extensions in the context of traffic flow 
have focused primarily on collective steady-state properties, captured by the fundamental diagram,
while the transient dynamics and relaxation processes 
as well as the trajectories of individual cars
have received relatively little attention.

In a recent paper, Jha et al. studied the 
transient evolution 
of CA rule 184
from a random initial configuration to steady-state conditions
\cite{Jha2025}. 
Unlike previous studies that treated the transient relaxation process as a mere prelude to stationarity, 
they analyzed the full space–time pattern of jams.
They introduced the notion of elementary jams, which act as the basic building 
blocks of these patterns, and used them to express key quantities such 
as the total delay and the relaxation time. 
They used finite-size scaling analysis of simulation results to calculate 
the distributions of jam sizes and lifetimes and the scaling of the relaxation time near the transition density.
This geometric approach 
reveals the statistical nature of the relaxation process 
and provides a new way to understand how CA rule 184 
evolves toward its steady-state.

In this paper, we pursue a more detailed analysis of the transient dynamics 
of the DSTF model by focusing on the statistical properties of the trajectories 
of individual cars.
We use the framework of first-passage processes 
\cite{Redner2001}
to derive a closed-form expression for
the distribution 
$P(T_{\rm FS}=t)$
of first-stopping (FS) times,
which is the probability that a randomly selected car will be stopped for the 
first time at time $t$.
We also obtain a closed-form expression for the 
stopping probability $P_{\rm S}(t)$,
which is the probability that a randomly selected car
will be stopped at time $t$.
In the low-density phase of $0 < p < 1/2$, 
the stopping probability $P_{\rm S}(t)$ yields a closed-form expression for  
the distribution 
$P(T_{\rm LS}=t)$
of last-stopping (LS) times,
which is the probability that a randomly selected car will be stopped for the
last time at time $t$, beyond which it will move freely indefinitely.
In this regime, we analyze the relation between the LS time and the 
number of stopping events $N_{\rm S}$ 
which take place up to that time.
We present closed-form expressions 
for the joint distribution $P(T_{\rm LS}=t,N_{\rm S}=n)$,
for the two conditional distributions that emanate from it
and for the marginal distribution $P(N_{\rm S}=n)$.
These results provide insight on the time scales of congestion
and relaxation in deterministic traffic flow.

The paper is organized as follows.
In Sec. II we review the DSTF model.
In Sec. III we derive a closed-form expression for the distribution $P(T_{\rm FS}=t)$
of first-stopping times and calculate its mean and variance. 
In Sec. IV we derive a closed-form expression for the probability $P_{\rm S}(t)$
that a randomly selected car will
be stopped at time $t$,
which is closely related to the mean collective velocity $v(t)$.
In Sec. V we present a closed-form expression for the distribution $P(T_{\rm LS}=t)$
of last-stopping times and calculate its mean and variance.
In Sec. VI we analyze the joint probability distribution 
$P(T_{\rm LS}=t,N_{\rm S}=n | N_{\rm S} \ge 1)$ of LS times and of the number of stopping events
that take place until that time,
and obtain closed-form expressions for the conditional distributions that 
emanate from this distribution.
The results are discussed in Sec. VII and summarized in Sec. VIII.
In Appendix A we derive a useful identity which is utilized in the
calculation of the stopping probability $P_{\rm S}(t)$.
In Appendix B we calculate the generating function of the 
distribution of LS times.

\section{The Model}

The DSTF model is defined on a finite 1D lattice
of length $L$ with periodic boundary conditions.
The lattice cells are labeled from left to right by $i=0,1,2,\dots,L-1$.
Each cell may be empty or occupied by a single car.
Starting from a random initial state with car density $p$, 
the system evolves via a parallel update rule, in which at every time
step each car moves one step to the right,
from cell $i$ to cell $i+1$ (mod $L$),
if the cell on its right is empty,
and is stopped if the cell on its right is occupied by another car.
In the rest of the paper, for simplicity, we drop the mod $L$ notation,
assuming that in all cases the modulo operation is taken whenever needed
in order to bring the index $i$ back into the range of $[0,L-1]$.
The state of the system at time $t$ can be described in terms of an array
$x_t(i)$, $i=0,1,\dots,L-1$,
where $x_t(i)=1$ if cell $i$ is occupied by a car at time $t$ 
and $x_t(i)=0$ if it is empty. 
In particular, the initial state at time $t=0$ is denoted by
$x_0(i)$, $i=0,1,\dots,L-1$.

\begin{figure}
\includegraphics[width=13.0cm]{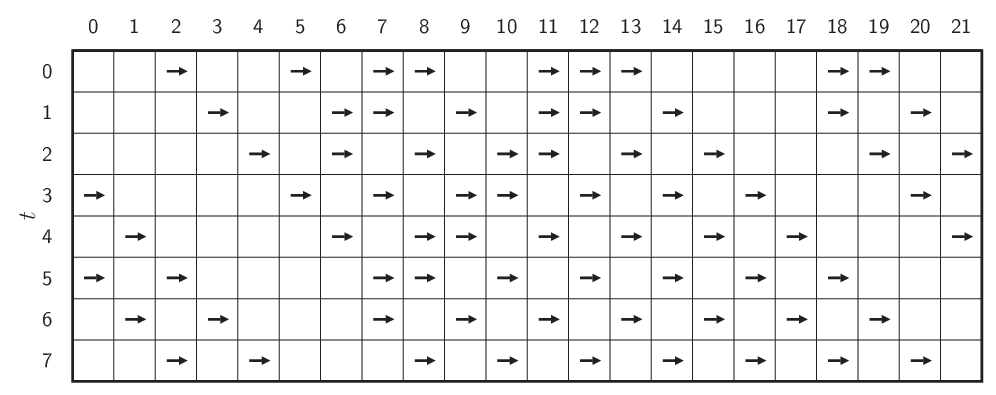} 
\caption{
Illustration of the dynamics of the DSTF model.
The horizontal coordinate represents the space dimension and the vertical coordinate
represents the time, which proceeds downward.
The model is defined on a 1D lattice of size $L=22$ with periodic boundary conditions.
In this example there are $9$ cars, where $4$ of them are stopped in the first time step.
The time evolution of the system is presented for $t=1,2,\dots,7$, where at $t=7$
it converges into an FFP state.
During the first six time steps, the cars that start at time $t=0$ from cells
$i=2,8,12$ and $18$ 
experience one stopping event, while
the cars that start from cells 
$i=5,7$ and $11$
experience two stopping events.
In contrast, the cars that start from cells 
$13$ and $19$ 
move freely right from
the beginning and are never stopped.
}
\label{fig:1}
\end{figure}

In Fig. \ref{fig:1} we present an
illustration of the dynamics of the DSTF model.
The horizontal coordinate represents the space dimension,
in which the cells are indexed by $i=0,\dots,L-1$
and $L=22$ is the lattice size.
The vertical coordinate
represents the time $t$, which proceeds downward, starting from the 
initial condition at $t=0$.
Cars are represented by right-pointing arrows and move to the right.
The boundary conditions are periodic, such that cars
that move out of cell $i=21$ enter cell $i=0$.
In this example there are $9$ cars, where $4$ of them are stopped at $t=1$.
The time evolution of the system is presented for $t=1,2,\dots,7$, where at $t=7$
it converges into a free-flowing periodic state.

In the DSTF model each car may interact directly only with the car in front of it
(on its right)
and the car behind it (on its left). 
However, the interaction is asymmetric. 
The motion of each car may be affected by the car in front
of it, but the car in front is not affected by the car behind.
This implies that as the cars flow downstream (to the right), the information
that affects their motion flows upstream (to the left).
It gives rise to indirect  
interactions which propagate backwards through the chain of cars.

The dynamics of the DSTF model coincides with CA rule 184 in 
Wolfram's numbering scheme 
\cite{Wolfram1983,Wolfram1984,Fuks2023}.
It also coincides with the deterministic limit of the Nagel-Schreckenberg
model with velocities $v=0,1$.
The setup of the model and the interactions between particles are the same as in the
totally asymmetric simple exclusion process (TASEP)
\cite{Spitzer1970,Derrida1992,Derrida1998},
which belongs to the Kardar-Parisi-Zhang universality class
\cite{Kardar1986}.
However, the dynamics differs in that the DSTF model 
evolves via a deterministic parallel update. While parallel-update 
variants of TASEP have been studied 
\cite{Schutz1993,Rajewsky1998}, 
they have primarily focused on collective or correlation properties rather 
than on trajectory-level first-passage observables.
In the hydrodynamic limit, CA rule 184 can be described by a
partial differential equation that captures the evolution of the
coarse-grained density of cars.
In this description, traffic jams appear as shocks which move
backwards as time evolves
\cite{Belitsky2011a,Belitsky2011b}.

\begin{figure}
\includegraphics[width=16.0cm]{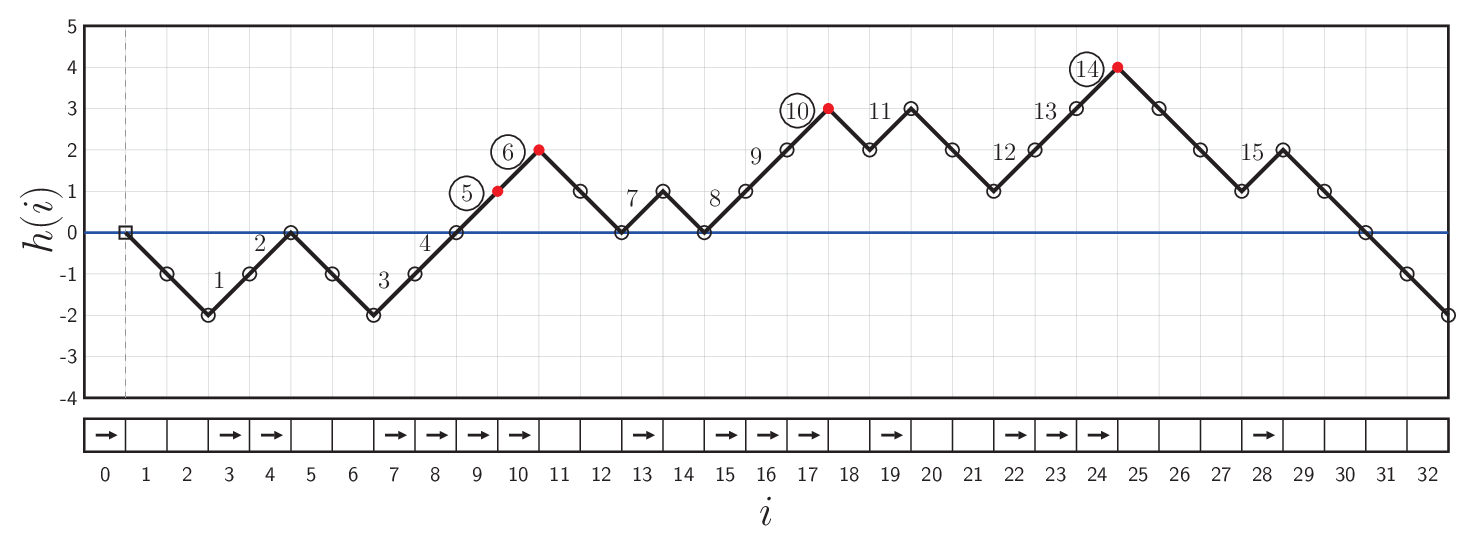} 
\caption{
Illustration showing an initial configuration of the 1D traffic model
at time $t=0$
and the corresponding mountain range.
The structure of the mountain range provides the stopping times of
the left-most car, which starts from cell $i=0$ at $t=0$,
referred to as the selected car.
In the mapping from the configuration of cars to the mountain range,
each occupied cell corresponds to an ascending step and each empty
cell corresponds to a descending step.
The time steps of the traffic model, which correspond to the ascending steps
of the mountain range, 
are listed from left to right
by $1,2,3,\dots$ along the ascending slopes.
The stopping times of the selected car 
(marked by circles)
correspond to the times at which
the mountain range reaches record heights
(red dots)
that exceed the heights that
have been reached in all the previous steps.
More specifically, the $k$th stopping time corresponds to the first time in
which the mountain range reaches a height of $h(i)=k$.
The last-stopping time corresponds to the last record height that is reached,
which is never exceeded in later time steps.
}
\label{fig:2}
\end{figure}

In Fig. \ref{fig:2} we present an
illustration of an initial configuration of cars at time $t=0$
in a segment of cells, which is a part of a larger system.
This segment is selected such that 
the leftmost cell is occupied by 
a randomly selected car.
The cells in the segment are indexed from left to right, 
where the leftmost cell is indexed by $i=0$.
The initial configuration 
$x_0(i)$, $i=1,2,\dots$,
of the cars  
located in front of the randomly selected car  
can be mapped into a discrete mountain landscape, 
where the positions of the cars determine the height profile.
Here we show the corresponding mountain landscape, in which 
each cell that is occupied by a car corresponds to an ascending step and each empty
cell corresponds to a descending step.
The reference height is determined by the leftmost
cell $i=0$, such that $h(0)=0$.
The height of the mountain landscape at the location of cell $i \ge 1$,
is given by

\begin{equation}
h(i) = \sum_{j=1}^{i} \left[ 2 x_0(j) - 1 \right].
\label{eq:h_i}
\end{equation}

\noindent
The structure of the mountain landscape captures the future dynamics 
of the randomly selected car,
initially located at $i=0$.
To extract this information, we index the ascending steps in
the mountain landscape from left to right by $t=1,2,\dots$,
while the descending steps are ignored. 
Some of the ascending steps establish a new height record, 
namely the height reached by these steps  
is higher than all the sites on their left-hand side.
For each time $t$, if the corresponding ascending step establishes
a new height record, the selected car will be stopped at time step $t$.
In contrast, if the corresponding ascending step does not establish
a new height record, the selected car will move freely at time step $t$.
Thus, the first-stopping time corresponds to the 
time $t$ associated with the ascending step
at which the mountain landscape reaches the height of $h(i)=1$
for the first time.
Similarly, the $k$th stopping time corresponds to the 
time $t$ associated with the ascending step at which the mountain landscape
reaches the height of $h(i)=k$ for the first time.
The last-stopping time corresponds to the 
time $t$ associated with the ascending step at which the mountain landscape
reaches the last record height,
which is never exceeded at later times.
Thus, there is a complete analogy between the pattern of stopping events
and the record statistics of 1D random walks
\cite{Mounaix2020,Majumdar2024}.

\begin{figure}
\includegraphics[width=16.0cm]{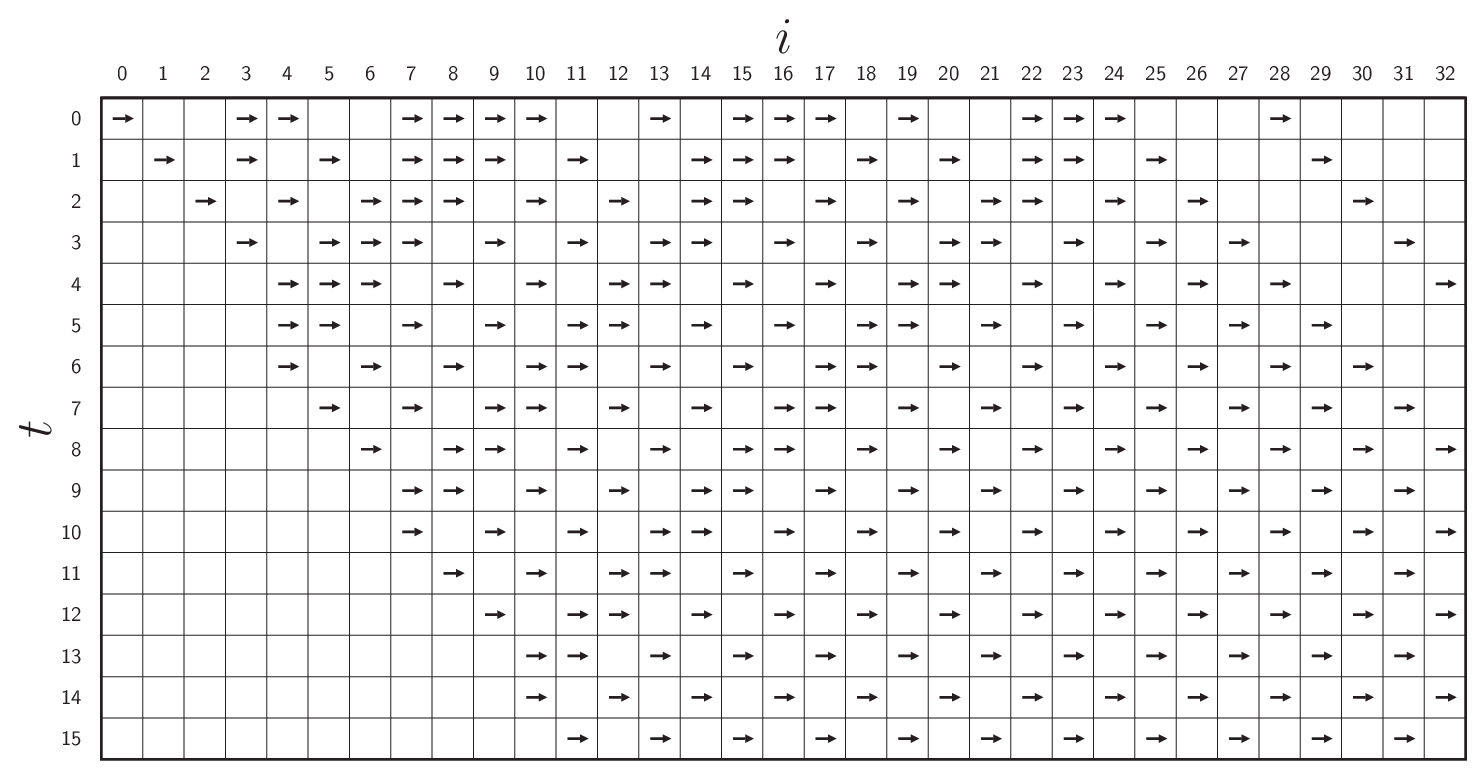} 
\caption{
The time evolution of the traffic flow that emerges from the initial
state shown in Fig. \ref{fig:2}.
This segment of 32 cells is a part of a larger system.
Thus, the boundaries of this subsystem are open and no
cars enter from the left within the time window that is presented.
The stopping time steps of the leftmost car, that starts from $i=0$
at time $t=0$ are $t=5, 6, 10$ and $14$.
These times coincide with the ascending records of the corresponding
mountain landscape, presented in Fig. \ref{fig:2},
where the records are marked by red dots and the times are
marked by circles.
The system converges to an FFP state at time $t=15$.
}
\label{fig:3}
\end{figure}

In Fig. \ref{fig:3} we present
the time evolution of the traffic flow that emerges from the initial
state shown in Fig. \ref{fig:2}.
This segment of 32 cells is a part of a larger system.
Thus, the boundaries of this subsystem are open.
For simplicity, we assume that no
cars enter from the left within the time window that is presented.
The stopping time steps of the leftmost car, that starts from $i=0$
at time $t=0$ are $t=5, 6, 10$ and $14$.
These times coincide with the ascending records of the corresponding
mountain landscape, presented in Fig. \ref{fig:2}.
The system converges to an FFP state at time $t=15$.

\section{The distribution of first-stopping times}

Consider a random initial state at time $t=0$ 
with car density $0 < p < 1$.
Starting at time $t=1$, at every time step 
each car moves to the next cell on the right unless it 
is occupied by another car.
Some of the cars are stopped at time $t=1$ and start moving only at
a later time (and may be stopped again later). 
Other cars start moving at time $t=1$ and are stopped
at later times.
In the case of $0 < p < 1/2$ there are cars that move freely
at all the time steps and are never stopped.

Below we calculate the distribution $P(T_{\rm FS}=t)$
of FS times, 
which is the probability that a randomly selected 
car will be stopped for the first time at time $t$, where $t \ge 1$.
For simplicity, we consider the infinite system limit,
where $L \rightarrow \infty$.
We select a random car at time $t=0$ and denote its
initial cell by $i=0$, such that $x_0(0)=1$. 
By inspection, it is found that the pattern of 
moving and stopping of this car during the first $t$ time steps is 
determined by the initial arrangement of cars in the $2t-1$ cells
in front of cell $i=0$. This arrangement is given by 
$x_0(1),x_0(2),\dots,x_0(2t-1)$.
In this framework, a necessary and sufficient condition for the car that initially resides in
cell $i=0$ to be stopped for the first time at time $t=1$ is $x_0(1)=1$.
The initial arrangements that yield a first-stopping event at time $t \ge 2$
for the car that initially resides in cell $i=0$
must satisfy the following three conditions:

\begin{enumerate}

\item
There is an equal number of occupied
and unoccupied cells in the first $2t-2$ cells in front of cell $i=0$,
namely $t-1$ cars and $t-1$ empty cells. 
This implies that $\sum\limits_{t'=1}^{2t-2} x_0(t') = t-1$.

\item
For any value of $1 \le j \le 2t-2$ the number of empty cells among the
first $j$ cells in front of cell $i=0$ is larger than or equal to
the number of cells occupied by cars.
This implies that
$\sum\limits_{t'=1}^{j} x_0(t') \le \lfloor j/2 \rfloor$,
where $\lfloor y \rfloor$ is the integer part of $y$.

\item
Cell $2t-1$ is occupied,
namely $x_0(2t-1)=1$.

\end{enumerate}

In the language of mountain ranges, the first condition means
that the mountain landscape returns to the reference
height at $i=2t-2$. This implies that not only $h(0)=0$
but also $h(2t-2)=0$.
The second condition means that at any point
within $1 \le j \le 2t-2$,
the height of the mountain landscape is non-positive, 
namely $h(j) \le 0$.
Mountain landscapes that satisfy these two conditions
are referred to as mountain ranges or Dyck paths
\cite{Flajolet2009}.
Note that in the common convention in the combinatorial literature,
the condition for a mountain range is $h(j) \ge 0$,
which is equivalent to the condition above by the symmetry $h \rightarrow -h$.

The number of possible configurations 
of the first $2t-1$ cells in front of cell $i=0$
that satisfy the three conditions specified above
is given by the Catalan number $C_{t-1}$,
defined by
\cite{Audibert2010,Stanley2015}

\begin{equation}
C_n = \frac{1}{n+1} \binom{2n}{n}.
\label{eq:Catalan}
\end{equation}

\noindent
The Catalan number $C_n$ appears in many combinatorial problems.
For example, it counts
the number of discrete mountain ranges
of length $2n$
\cite{Audibert2010,Stanley2015,Koshy2009,Deutsch1999}.
The Catalan number $C_{n-1}$
counts the number of first-return trajectories 
of a random walk on a one-dimensional lattice 
that return to the origin for the first time after $2n$ time steps
\cite{Gruda2025}.

The Catalan number can also be expressed in the form
\cite{Koshy2009,Stanley2015}

\begin{equation}
C_{ n } = 
\binom{2n}{n} - \binom{2n}{n+1}.
\label{eq:Catalan3}
\end{equation}

\noindent
The expression of Eq. (\ref{eq:Catalan3})  
clearly shows that the Catalan number must be an integer.
The first term on the right-hand side of Eq. (\ref{eq:Catalan3})
is a central binomial coefficient that counts all the mountain landscapes
that return to the reference height at $i=2n$. The second term on the 
right-hand side of Eq. (\ref{eq:Catalan3}) counts the mountain landscapes that
cross the reference height
from the positive side to the negative side or vice versa  
within the range of $0 < i <2n$.
Thus, the difference between these two terms accounts for the number of
mountain landscapes that return to the reference height at $i=2n$ without crossing
within the range of $0 < i < 2n$
(note that mountain landscapes that return to the reference height
but do not cross it are included).

Using the considerations presented above,
it is found that the distribution of FS times is given by

\begin{equation}
P(T_{\rm FS}=t) = C_{t-1} p^t (1-p)^{t-1},
\label{eq:PTFSt}
\end{equation}
 
\noindent
where $t \ge 1$.
The Catalan number on the right-hand side of Eq. (\ref{eq:PTFSt})
counts the number of possible configurations in front of a randomly selected 
car for which the FS time is $t$, while the powers of $p$ and $1-p$ account
for the probability of each one of these configurations to appear.

The generating function of the distribution $P(T_{\rm FS}=t)$ 
is given by

\begin{equation}
G(z) = \sum_{t=1}^{\infty}   z^t P(T_{\rm FS}=t).
\label{eq:Gx}
\end{equation}
 
\noindent
In the special case of $ z = 1$, the generating function $G( z )$ is reduced to

\begin{equation}
G(1)=\sum_{t=1}^{\infty} P(T_{\rm FS}=t),
\label{eq:G1}
\end{equation}

\noindent
which is the overall probability that a car selected randomly  at time $t=0$
will be stopped at least once.
Eq. (\ref{eq:G1}) can also be written in the form

\begin{equation}
G(1) = P(T_{\rm FS} < \infty).
\label{eq:G1PTFSinf}
\end{equation}

\noindent
The complementary probability of non-stopping (NS) cars,  
namely cars that will never be stopped, 
is given by

\begin{equation}
P_{\rm NS} = 1 - P(T_{\rm FS} < \infty).
\label{eq:PNS}
\end{equation}

\noindent
Inserting $P(T_{\rm FS}=t)$ from Eq. (\ref{eq:PTFSt}) into Eq. (\ref{eq:Gx})
and carrying out the summation, we obtain a closed-form expression for 
the generating function, which is given by

\begin{equation}
G( z ) = \frac{ 1 - \sqrt{ 1 - 4p(1-p)   z  } }{2(1-p)}.
\label{eq:Gx2}
\end{equation} 

\noindent
Inserting $  z = 1$ in Eq. (\ref{eq:Gx2})
and using Eq. (\ref{eq:G1PTFSinf}), 
we obtain

\begin{equation}
P(T_{\rm FS} < \infty) = \frac{ 1 - \sqrt{ (2p-1)^2 }}{2(1-p)}.
\label{eq:Gx3}
\end{equation} 

\noindent
Simplifying the right-hand side of Eq. (\ref{eq:Gx3}) 
for $0 < p < 1/2$, we obtain

\begin{equation}
P(T_{\rm FS} < \infty)  =
\frac{p}{1-p}.
\label{eq:PTFSinfl12}
\end{equation}

\noindent
Inserting $P(T_{\rm FS} < \infty)$ 
from Eq. (\ref{eq:PTFSinfl12})
into the right-hand side of Eq. (\ref{eq:PNS}),
it is found that in the low-density phase, where $0 < p < 1/2$,
the probability that a randomly 
selected car will never be stopped is given by

\begin{equation}
P_{\rm NS} = 
\frac{ 1 - 2p }{1 - p}.
\label{eq:PNS1}
\end{equation}

\noindent
In the low-density limit of $p \rightarrow 0$ 
it is found that $P_{\rm NS} \rightarrow 1$,
namely most of the cars move freely 
without obstruction right from the beginning.
As the car density is increased 
the fraction of cars that are never stopped 
decreases until it vanishes in the limit of
$p \rightarrow 1/2^{-}$.

Simplifying the right-hand side of Eq. (\ref{eq:Gx3}) 
for $1/2 < p < 1$, it yields
$P(T_{\rm FS} < \infty)  = 1$.
Thus, in this regime the distribution of FS times,
given by
Eq. (\ref{eq:PTFSt}), 
is normalized.
This implies that in the high-density phase every single car
is stopped at least once. 
Moreover, after its FS event, each car undergoes an 
infinite sequence of further stopping events.

In order to obtain a normalized distribution of FS times for $0 < p < 1/2$, 
we condition on  
cars that experience at least one stopping event.
The conditional distribution of FS times is given by

\begin{equation}
P(T_{\rm FS}=t | T_{\rm FS} < \infty) = \frac{ P(T_{\rm FS}=t) }{ P(T_{\rm FS} < \infty) }.
\label{eq:PTFStTFSinf}
\end{equation}

\noindent
Inserting 
$P(T_{\rm FS}=t)$
from Eq. (\ref{eq:PTFSt}) and
$P(T_{\rm FS} < \infty)$ 
from Eq. (\ref{eq:PTFSinfl12})
into the right-hand side of Eq. (\ref{eq:PTFStTFSinf}), 
we obtain a closed-form expression for the conditional distribution of FS times
in the low-density phase,
which is given by

\begin{equation}
P(T_{\rm FS}=t | T_{\rm FS} < \infty) = C_{t-1} p^{t-1} (1-p)^t, 
\label{eq:PTFStTFSinf2}
\end{equation}

\noindent
where the Catalan number $C_{t-1}$ is given by Eq. (\ref{eq:Catalan}) and $t \ge 1$.

In the special case of $p=1/2$,
namely at the phase transition,
the distribution of FS times is reduced to

\begin{equation}
P(T_{\rm FS}=t) =   C_{t-1}  \left( \frac{1}{2} \right)^{2t-1},
\label{eq:PTFSt12}
\end{equation}

\noindent
where $t \ge 1$.

The distribution $P(T_{\rm FS}=t)$, given by Eq. (\ref{eq:PTFSt}),
is analogous to the distribution of first-return (FR) times of a random walk (RW)
on a 1D lattice, which at each time step
moves to the right with probability $1-p$ and
to the left with probability $p$
\cite{Klinger2022,Gruda2025}.
Starting from the origin at time $t=0$ and
moving to the right at $t=1$, the probability that 
such RW will return to the origin for the first 
time at time $t=2n$ is given by

\begin{equation}
P(T_{\rm FR}=2n) = C_{n-1} p^n (1-p)^{n-1}.
\label{eq:PTFR2n}
\end{equation}

\noindent
Comparing Eqs. (\ref{eq:PTFSt}) and (\ref{eq:PTFR2n}),
it is found that
$P(T_{\rm FS}=t) = P(T_{\rm FR}=2t)$.
For $p \ge 1/2$ the RW is recurrent
\cite{Polya1921},
namely the probability that
it will eventually return to the origin is $P_{\rm R}=1$.
In contrast, for $p < 1/2$ the RW is transient
\cite{Polya1921}, 
namely the probability
that it will ever return to the origin is only

\begin{equation}
P_{\rm R} = \frac{p}{1-p},
\end{equation}
 
\noindent
which is analogous to $P(T_{\rm FS} < \infty)$,
given by Eq. (\ref{eq:PTFSinfl12}).

To analyze the long time limit of the distribution of FS times,
we invoke the asymptotic form of the Catalan number at $n \gg 1$,
which is given by
\cite{Olver2010}

\begin{equation}
C_n = \frac{ 4^n }{ \sqrt{\pi} n^{3/2} } + {\mathcal O} \left( \frac{1}{n^{5/2}} \right).
\label{eq:Catalan_tail}
\end{equation}

\noindent
Using this result, it is found that  
for $0 < p < 1/2$
the tail of the conditional distribution of FS times 
follows an exponentially truncated power-law decay of the form

\begin{equation}
P(T_{\rm FS}=t | T_{\rm FS} < \infty) \simeq \frac{1-p}{ \sqrt{\pi}  } \frac{ [4p(1-p)]^{t-1} }{ t^{3/2} }.
\label{eq:PTFStTFSinf3}
\end{equation}

\noindent
The decay is fastest in the dilute system limit of $p \rightarrow 0$
and becomes slower as $p$ is increased towards the transition density
of $p=1/2$.
A similar result is found for $1/2 < p < 1$, where the tail of the distribution
of FS times is given by

\begin{equation}
P(T_{\rm FS}=t) \simeq \frac{p}{ \sqrt{\pi} } \frac{ [4p(1-p)]^{t-1} }{ t^{3/2} }.
\label{eq:PTFStaspt}
\end{equation}

\noindent
Note that the expressions on the
right-hand sides of Eqs. (\ref{eq:PTFStTFSinf3}) and  (\ref{eq:PTFStaspt})
are symmetric with respect to $p=1/2$.
In the special case of $p=1/2$ it is found that in the long-time limit

\begin{equation}
P(T_{\rm FS}=t) \simeq \frac{1}{ 2 \sqrt{\pi} }   \frac{1}{ t^{3/2} },
\label{eq:PTFStaspt12}
\end{equation}

\noindent
namely the tail of the distribution of FS times follows a pure power-law decay,
which is typical of continuous phase transitions.

As discussed above, the mapping of the traffic dynamics to a mountain landscape
implies that stopping events correspond to ascending records of the 
corresponding one-dimensional random walk, also referred to as 
ascending ladder epochs
\cite{Feller1971a,Feller1971b,Spitzer1970,Majumdar2024}.
A key property of ladder epochs is that they are regeneration points,
which means that once a new record height is reached, the future
evolution is statistically independent of the past.
Consequently, the durations of periods of uninterrupted motion of a car
between successive stopping events form a renewal process.
These durations are thus independent and identically distributed, following
the distribution of FS times.
In other words, once a stopping event has taken place, 
the waiting time for the next stopping event follows the
same statistics as the first-stopping event.
The waiting time between successive records is referred to as the
record age
\cite{Majumdar2012,Mounaix2020}.

\begin{figure}
\includegraphics[width=8.1cm]{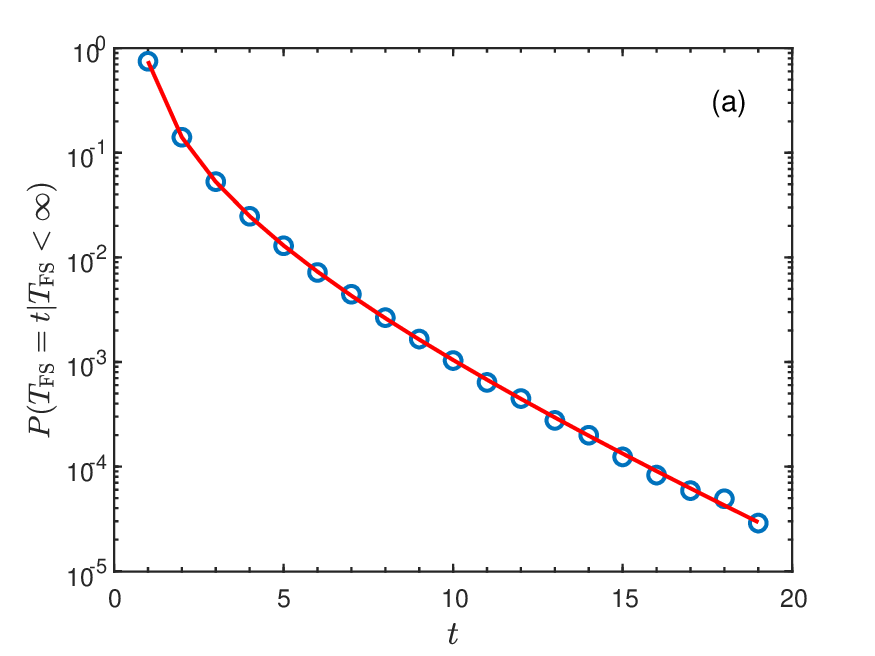} 
\includegraphics[width=8.1cm]{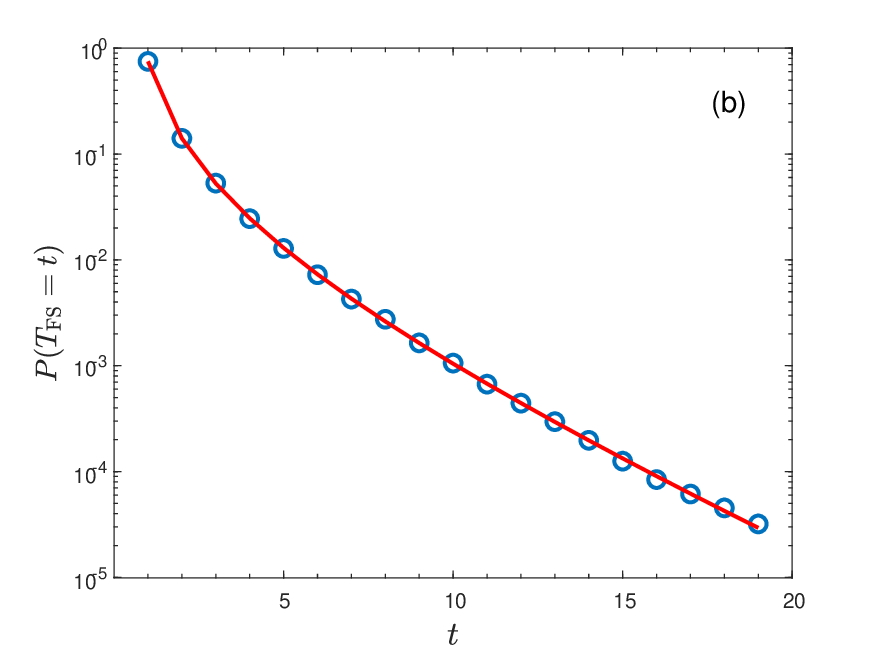} 
\caption{
(a) Analytical results (solid line),
obtained from Eq. (\ref{eq:PTFStTFSinf2}),
for the distribution 
$P(T_{\rm FS}=t | T_{\rm FS} < \infty)$
of FS times 
in the DSTF model
for random initial states with car density
of $p=0.25$. 
(b) Analytical results (solid line),
obtained from Eq. (\ref{eq:PTFSt}),
for the distribution
$P(T_{\rm FS}=t)$ 
of FS times for random initial states with 
car density of $p=0.75$.
The analytical results are in very good agreement with the results obtained from
computer simulations (circles) on a lattice of size $L=10,000$.
Note that for any value of $p>1/2$,
the results for $P(T_{\rm FS}=t)$,
obtained from Eq. (\ref{eq:PTFSt})
coincide with the results for
$P(T_{\rm FS}=t | T_{\rm FS} < \infty)$,
evaluated at $p^{\prime} = 1 - p$.
}
\label{fig:4}
\end{figure}

In Fig. \ref{fig:4} we present 
the distribution of FS times
in the low-density and high-density phases
of the DSTF model.
In Fig. \ref{fig:4}(a) we present
analytical results (solid line),
obtained from Eq. (\ref{eq:PTFStTFSinf2}),
for the conditional distribution 
$P(T_{\rm FS}=t | T_{\rm FS} < \infty)$
of FS times for random initial states with car density
of $p=0.25$.
In Fig. \ref{fig:4}(b) we present
analytical results (solid line),
obtained from Eq. (\ref{eq:PTFSt}),
for the distribution 
$P(T_{\rm FS}=t)$
of FS times for random initial states with car density
of $p=0.75$.
The analytical results are in very good agreement with the results obtained from
computer simulations (circles) on a lattice of length $L=10,000$.
The simulation results for $p=0.25$ ($p=0.75$) are averaged over 1,000 (100) initial
configurations.
Note that for any value of $1/2 < p < 1$ 
the results for $P(T_{\rm FS}=t)$,
obtained from Eq. (\ref{eq:PTFSt}),
coincide with the results for
$P(T_{\rm FS}=t | T_{\rm FS} < \infty)$,
obtained from Eq. (\ref{eq:PTFStTFSinf2}) 
with $p^{\prime} = 1 - p$.
Indeed, these distributions display
an initial curved (power-law–like) regime at short times
and then cross over to an asymptotic exponential tail, 
characteristic of exponentially-truncated power-law distributions.

\begin{figure}
\includegraphics[width=9.0cm]{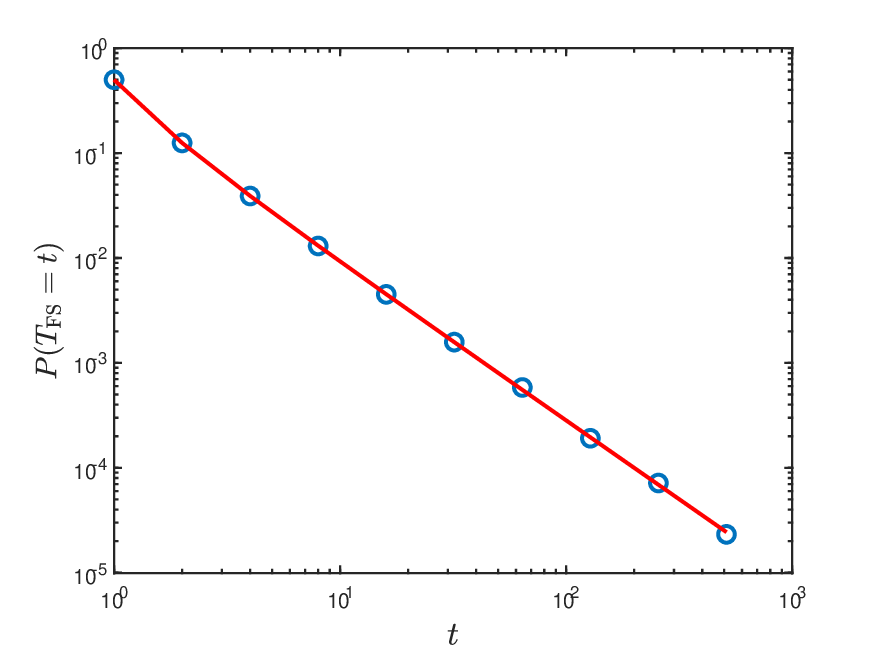} 
\caption{
Analytical results (solid line),
obtained from Eq. (\ref{eq:PTFSt12}),
for the distribution 
$P(T_{\rm FS}=t)$
of FS times 
in the DSTF model
for random initial states 
in the special case in which the car density
is $p=0.5$. 
The analytical results are in very good agreement with the results obtained from
computer simulations (circles) on a lattice of size $L=10,000$.
}
\label{fig:5}
\end{figure}

In Fig. \ref{fig:5} we present
analytical results (solid line),
obtained from Eq. (\ref{eq:PTFSt12}),
for the distribution
$P(T_{\rm FS}=t)$ 
of FS times in the DSTF model 
for random initial states, in the special case in which the
car density is $p=1/2$, namely at the phase transition.
The analytical results are in very good agreement with the results obtained from
computer simulations (circles) on a lattice of size $L=10,000$.
The simulation results are averaged over 500 initial configurations.
In this case
$P(T_{\rm FS}=t)$ 
follows a power-law distribution at all time scales
and is not truncated by an exponential tail.

The $r$th moment of the conditional distribution of 
FS times in the low-density phase is given by

\begin{equation}
{\mathbb E}[T_{\rm FS}^r | T_{\rm FS} < \infty] =
\sum_{t=1}^{\infty}
t^r
P(T_{\rm FS}=t | T_{\rm FS} < \infty).
\label{eq:ETFSr}
\end{equation}
 
\noindent
Inserting 
$P(T_{\rm FS}=t | T_{\rm FS} < \infty)$
from Eq. (\ref{eq:PTFStTFSinf2}) into Eq. (\ref{eq:ETFSr}),
setting $r=1$ and carrying out the summation,
it is found that the mean FS time is

\begin{equation}
{\mathbb E}[T_{\rm FS} | T_{\rm FS} < \infty] = \frac{ 1 - p }{ 1 - 2p }.
\label{eq:ETFSTFSinf}
\end{equation}

\noindent
In the low-density limit of $p \rightarrow 0$, the mean FS time satisfies
${\mathbb E}[T_{\rm FS} | T_{\rm FS} < \infty] \rightarrow 1$.
This implies that in the low-density limit most of the FS 
events take place at time $t=1$.
Furthermore, the mean FS time increases monotonically as $p$ is increased
and diverges at $p \rightarrow 1/2^{-}$.
 
Inserting 
$P(T_{\rm FS}=t | T_{\rm FS} < \infty)$
from Eq. (\ref{eq:PTFStTFSinf2}) into Eq. (\ref{eq:ETFSr}),
setting $r=2$ and carrying out the summation,
it is found that  
the second moment of 
$P(T_{\rm FS}=t | T_{\rm FS} < \infty)$
is given by

\begin{equation}
{\mathbb E}[T_{\rm FS}^2 | T_{\rm FS} < \infty] = \frac{ (1 - p)(1-2p+2p^2) }{ (1 - 2p)^3 }.
\label{eq:ETFS2inf}
\end{equation}

\noindent
The variance of 
$P(T_{\rm FS}=t | T_{\rm FS} < \infty)$
is given by

\begin{equation}
{\rm Var}(T_{\rm FS} | T_{\rm FS} < \infty) 
=
{\mathbb E}[T_{\rm FS}^2 | T_{\rm FS} < \infty]
-
{\mathbb E}[T_{\rm FS} | T_{\rm FS} < \infty]^2.
\label{eq:VarTFS}
\end{equation}

\noindent
Inserting
${\mathbb E}[T_{\rm FS} | T_{\rm FS} < \infty]$
from Eq. (\ref{eq:ETFSTFSinf})
and
${\mathbb E}[T_{\rm FS}^2 | T_{\rm FS} < \infty]$
from Eq. (\ref{eq:ETFS2inf})
into Eq. (\ref{eq:VarTFS})
and rearranging terms,
we obtain

\begin{equation}
{\rm Var}(T_{\rm FS} | T_{\rm FS} < \infty) = \frac{ p(1-p) }{(1-2p)^3}.
\label{eq:VarTFSTFSinf}
\end{equation}

\noindent
The variance vanishes in the low-density limit of $p \rightarrow 0$ and
diverges at $p \rightarrow 1/2^{-}$.

In the high-density phase, where $1/2 < p < 1$, 
the distribution of FS times is given by
Eq. (\ref{eq:PTFSt}).
As mentioned above, for
$1/2 < p < 1$, we set
$P(T_{\rm FS} < \infty) =1$,
which implies that the distribution $P(T_{\rm FS}=t)$, given by  
Eq. (\ref{eq:PTFSt})
is normalized.
The $r$th moment of $P(T_{\rm FS} = t)$ is given by

\begin{equation}
{\mathbb E}[T_{\rm FS}^r] = 
\sum_{t=1}^{\infty}
t^r P(T_{\rm FS} = t).
\label{eq:ETFSrpg12}
\end{equation}

\noindent
Inserting $P(T_{\rm FS} = t)$ from Eq. (\ref{eq:PTFSt})
into Eq. (\ref{eq:ETFSrpg12}), setting $r=1$
and carrying out the summation,
it is found that
the mean FS time for $1/2 < p < 1$ is given by 

\begin{equation}
{\mathbb E}[T_{\rm FS}] = \frac{p}{2p-1}.
\label{eq:ETFSpg12}
\end{equation}

\noindent
It diverges at $p \rightarrow 1/2^{+}$ and
decreases monotonically as $p$ is increased.
Comparing Eqs. (\ref{eq:ETFSTFSinf}) and (\ref{eq:ETFSpg12}),
it is found that the mean FS time (for cars that are stopped at least once)
is symmetric around $p=1/2$.

Inserting $P(T_{\rm FS} = t)$ from Eq. (\ref{eq:PTFSt})
into Eq. (\ref{eq:ETFSrpg12}), setting $r=2$ and carrying out the summation,
it is found that
the second moment of the distribution of FS times is given by

\begin{equation}
{\mathbb E}[T_{\rm FS}^2] = \frac{p(1-2p+2p^2)}{(2p-1)^3}.
\label{eq:ETFS2pg12}
\end{equation}

\noindent
Thus, the variance is given by

\begin{equation}
{\rm Var}(T_{\rm FS}) = \frac{ p(1-p) }{ (2p-1)^3 }.
\label{eq:VarTFSpg12}
\end{equation}

\noindent
The variance diverges at $p \rightarrow 1/2^{+}$ and decreases
monotonically as $p$ is increased in the range of $1/2 < p < 1$.
Comparing Eqs. (\ref{eq:VarTFSTFSinf}) and (\ref{eq:VarTFSpg12}) it is found that the variance
of the distribution of FS times (for cars that are stopped at least once)
is symmetric around $p=1/2$.

\begin{figure}
\includegraphics[width=11.0cm]{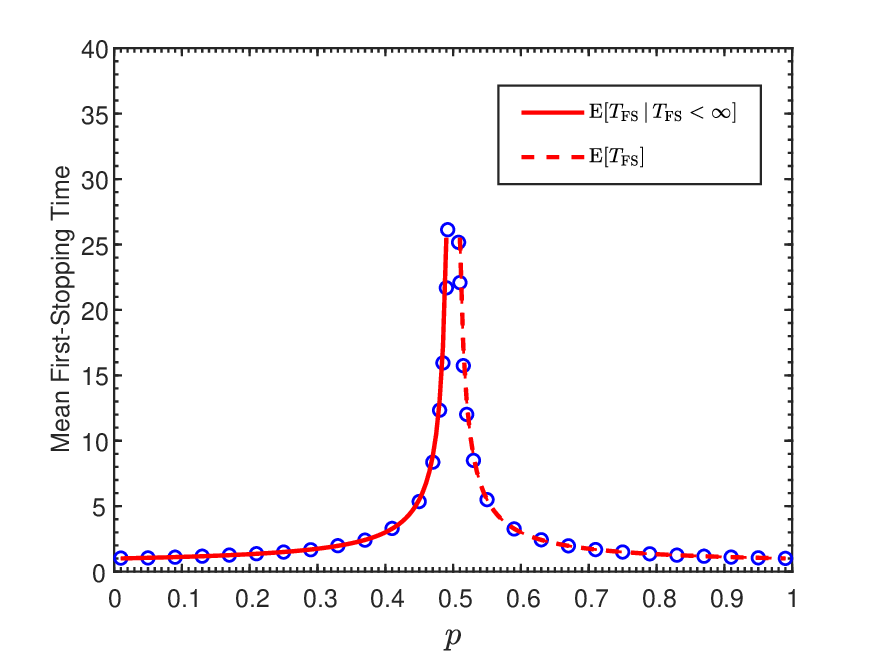} 
\caption{
Solid line: Analytical results
obtained from Eq. (\ref{eq:ETFSTFSinf}),
for the mean FS times
${\mathbb E}[T_{\rm FS} | T_{\rm FS} < \infty]$
in the low-density phase of the DSTF model,
conditioned on cars that are stopped at least once,
as a function of the car density $0 < p < 1/2$.
Dashed line: Analytical results,
obtained from Eq. (\ref{eq:ETFSpg12}),
for the mean FS times
${\mathbb E}[T_{\rm FS}]$
in the high-density phase of the DSTF model,
as a function of the car density $1/2 < p < 1$.
The analytical results are
in very good agreement with the results obtained from 
computer simulations for a system of size of $L=10,000$ (circles).
}
\label{fig:6}
\end{figure}

In Fig. \ref{fig:6} we present
analytical results (solid line),
obtained from Eq. (\ref{eq:ETFSTFSinf}),
for the mean FS time
${\mathbb E}[T_{\rm FS} | T_{\rm FS} < \infty]$
in the DSTF model 
as a function of the car density 
for $0 < p < 1/2$,
conditioned on cars that are stopped at least once.
We also present analytical results (dashed line),
obtained from Eq. (\ref{eq:ETFSpg12}),
for the mean FS time
${\mathbb E}[T_{\rm FS}]$
as a function of the car density for $1/2 < p < 1$.
The analytical results are
in very good agreement with the results obtained from 
computer simulations (circles) for a system of size $L=10,000$.
The simulation results are averaged over 50 initial configurations
for each value of $p$.

\section{The time-dependence of the stopping probability}

Following their first-stopping event, cars may be stopped again at later times.
For $0 < p < 1/2$ each car may be stopped only a finite number of times,
while for $1/2 < p < 1$ each car is stopped infinitely many times.
Below we calculate the probability $P_{\rm S}(t)$ that a randomly selected
car will be stopped at a given time $t$, where $t \ge 1$.

Consider a randomly selected car at time $t=0$ and denote its initial cell by $i=0$.
The question of whether this car will move or be stopped at time step $t$ is 
determined by the initial arrangement of the first $t$ cars in front of 
cell $i=0$.
The total number of empty cells that lie between these $t$ cars
is denoted by $\ell$, such that 
the initial state of the first $t+\ell$ cells in front of cell $i=0$ is denoted
by $x_0(1),x_0(2),\dots,x_0(t+\ell)$. 
In this framework, a necessary and sufficient condition for the car that resides at
cell $i=0$ to be stopped at time $t=1$ is $x_0(1)=1$.
The arrangements in which the car that resides at cell $i=0$ will be stopped at time $t \ge 2$
must satisfy the following three conditions:

\begin{enumerate}

\item
The number of empty cells in the interval that includes the first $t$ cars in front of cell $i=0$
is in the range of $0 \le \ell \le t-1$.
This implies that
in the  interval that includes the first $t$ cars in front of cell $i=0$ there are
more cars than empty cells.

\item
For any value of $1 \le j \le t + \ell - 1$
the number of occupied cells in the interval
$t+\ell-j, t+\ell-j+1,\dots,t+\ell-1$
is larger than or equal to the number of empty cells,
namely
$\sum\limits_{j'=1}^{j} x_0(t+\ell-j') \ge \lfloor (j+1)/2 \rfloor$.

\item
Cell $t+\ell$ is occupied,
namely $x_0(t+\ell)=1$.

\end{enumerate}

The number of initial configurations that satisfy these
three conditions is given by the Lobb number
$L_{\frac{t-1+\ell}{2},\frac{t-1-\ell}{2}}$,
which is defined by
\cite{Lobb1999,Koshy2012}

\begin{equation}
L_{n,m} = \frac{2m+1}{n+m+1} \binom{2n}{m+n}.
\label{eq:Lobb}
\end{equation}

\noindent
Using these results,
it is found that the probability that a randomly selected car
will be stopped at time $t$ is given by

\begin{equation}
P_{\rm S}(t) = \sum_{\ell=0}^{t-1} L_{ \frac{t-1+\ell}{2},\frac{t-1-\ell}{2} } p^t (1-p)^{\ell}.
\label{eq:PSt}
\end{equation}
 
\noindent
The Lobb number appears in many combinatorial problems
\cite{Lobb1999,Koshy2012}.
For example, $L_{n,m}$ counts the number of ways in which 
$n+m$ open parentheses and $n-m$ close parentheses
can be arranged to form the beginning of a valid sequence of balanced parentheses.
The Lobb number is a generalization of the Catalan number, 
which counts the number of complete valid sequences of balanced parentheses of a given length.

Inserting the Lobb number from Eq. (\ref{eq:Lobb}) into Eq. (\ref{eq:PSt}),
we obtain

\begin{equation}
P_{\rm S}(t) = \frac{p^t}{t} \sum_{\ell=0}^{t-1} (t-\ell) \binom{t-1+\ell}{t-1} (1-p)^{\ell}.
\label{eq:PSt2}
\end{equation}

\noindent
Expressing the right-hand side of Eq. (\ref{eq:PSt2}) in terms of a hypergeometric function,
the probability $P_{\rm S}(t)$ for $0 < p <1$ can be 
written in the form

\begin{equation}
P_{\rm S}(t) = 
\frac{2p - 1}{p}  
+ 
\binom{2t}{t-1} 
\frac{ [p(1-p)]^{t+1} }{pt}
\, _2 F_1
\left(\left. \begin{array}{c}
2,2t+1  \\
t+2 
\end{array}
\right|
 1-p  \right).
\label{eq:PSt_hg}
\end{equation}

\noindent
The function
$_2F_1( \ )$ on the right-hand side of Eq. (\ref{eq:PSt_hg})
is the hypergeometric function,
which is given by
\cite{Olver2010}

\begin{equation}
_2F_1 \left( \left.
\begin{array}{c}
a, b \\
c
\end{array}
\right| z 
\right) =
\sum_{n=0}^{\infty} 
\frac{ (a)_n (b)_n }{ (c)_n } \frac{ z^n }{ n! },
\label{eq:2F1}
\end{equation}

\noindent
where $(q)_n$ is the (rising) Pochhammer symbol
\cite{Olver2010}.
Note that the hypergeometric function is invariant to 
exchanging the parameters $a$ and $b$.

While
Eq. (\ref{eq:PSt_hg}) 
is valid in the whole range of $0 < p < 1$,
it is particularly convenient in the high-density phase, where $1/2 < p < 1$.
This is due to the fact that the asymptotic value of $(2p-1)/p$
appears explicitly, while the second term on the right-hand side
decays in the long time limit.
In the low-density phase of $0 < p < 1/2$
it is advantageous to use an equivalent expression,
given by

\begin{equation}
P_{\rm S}(t) = 
\binom{2t}{t-1} 
\frac{ [p(1-p)]^{t+1} }{p t}
\, _2 F_1
\left(\left. \begin{array}{c}
2,2 t+1 \\
t + 2
\end{array}
\right|
p  \right).
\label{eq:PSt_hg2}
\end{equation}

\noindent
The equivalence between the right-hand sides of
Eqs. (\ref{eq:PSt_hg})
and (\ref{eq:PSt_hg2}) 
is shown in Appendix A.

Since $P_{\rm S}(t)$ is the probability that a randomly selected car will be stopped
at time $t$ and $1-P_{\rm S}(t)$ is the probability that it will move,
the average velocity of cars at time $t$ is given  by 

\begin{equation}
\langle v \rangle_t = 1 - P_{\rm S}(t).
\end{equation}

\noindent
Using Eq. (\ref{eq:PSt_hg2}),
we express the average velocity at time $t$ for $0 < p < 1/2$ by

\begin{equation}
\langle v \rangle_t = 
1 -
\binom{2t}{t-1} 
\frac{ [p(1-p)]^{t+1} }{p t}
\, _2 F_1
\left(\left. \begin{array}{c}
2,2 t+1 \\
t + 2
\end{array}
\right|
p  \right).
\label{eq:PSt_hg2v} 
\end{equation}

\noindent
Similarly, using Eq. (\ref{eq:PSt_hg}),
we express the average velocity at time $t$ for $1/2 < p < 1$ by

\begin{equation}
\langle v \rangle_t = 
\frac{1 - p}{p}  
- 
\binom{2t}{t-1} 
\frac{ [p(1-p)]^{t+1} }{pt}
\, _2 F_1
\left(\left. \begin{array}{c}
2,2t+1  \\
t+2 
\end{array}
\right|
 1-p  \right).
\label{eq:PSt_hgv}
\end{equation}

\noindent
In both cases the second term on the right-hand side decays as time evolves.

To examine the behavior at the transition point, we insert
$p=1/2$ in Eq. (\ref{eq:PSt_hg}) and obtain

\begin{equation}
P_{\rm S}(t) = \frac{t+1}{t 4^t} \binom{2t}{t-1}.
\label{eq:PSt12}
\end{equation}

\noindent
In the long time limit, Eq. (\ref{eq:PSt12}) can be approximated by

\begin{equation}
P_{\rm S}(t) 
\simeq 
\frac{1}{\sqrt{\pi t}} 
+  \mathcal{O} \left( \frac{1}{t^{3/2}} \right).
\label{eq:PStaprx}
\end{equation}

Such power-law decays are common in coarsening processes that
are driven by the interaction between pairs of particles
\cite{Toussaint1983,Kang1984,Bramson1988,Krapivsky2010}.
In unary systems that consist of a single species of diffusive particles,
which undergo a pair annihilation reaction of the form 
$A + A \rightarrow \phi$,
the power-law decay may last for many orders of magnitude
\cite{Toussaint1983}.
In the case of binary systems that consist of two species of diffusive particles,
which undergo a pair annihilation reaction of the form $A + B \rightarrow \phi$,
the situation is more complicated.
If the densities of the two species are not the same,
the power-law decay is eventually truncated,
such that the density of the majority species saturates
while the density of the minority species follows an exponential tail 
\cite{Kang1984,Bramson1988}.

To cast the DSTF model in the framework of coarsening processes,
we note that the cars ($\rightarrow$) and vacancies ($\square$)
are conserved quantities which do not annihilate. Thus, the simplest
non-conserved objects which may annihilate are pairs of successive cars
($A = \rightarrow  \rightarrow$) and pairs of successive vacancies
($B = \square  \square$).
These pairs may be considered as quasi-particles that reside on the edges that
connect successive cells in the lattice. These edges form a 1D lattice
which is the dual lattice of the  lattice of cells.
Apart from the $A$ and $B$ quasi-particles, there are edges that
connect occupied and vacant sites of the form $\phi = \rightarrow \square$
(or $\square \rightarrow$).
These edges form the vacuum cells of the dual lattice that surround the
$A$ and $B$ quasi-particles.
As time evolves, the $A$ particles move ballistically to the left, 
reflecting the fact that traffic jams move backwards.
In contrast, the $B$ particles move ballistically to the right.
Pairs of $A$ and $B$ particles annihilate upon encounter.
This analogy casts the time evolution of the DSTF model in the framework of
ballistic annihilation in one dimension
\cite{Belitsky2005}.
The problem of ballistic annihilation in a continuous 1D
space was studied in Ref. 
\cite{Elskens1985}.
In the initial state there is a density $\sigma$ of particles
which are randomly distributed. 
The particles move ballistically such that the velocity of each particle
is $v$ with probability $p$ and $-v$ with probability $1-p$.
Pairs of particles that move in opposite directions are annihilated upon 
encounter.
For $p=1/2$ it was found that the density of particles at time $t$
scales like $\rho(t) \propto t^{-1/2}$.
In contrast for $p \ne 1/2$, the density of the minority population scales like
$\rho(t) \propto t^{-3/2} e^{- \alpha t}$,
where the parameter $\alpha$ is a function of the density $\sigma$,
the velocity $v$ and the probability $p$.
The DSTF model is analogous to the ballistic annihilation process on a 1D lattice.
The stopping probability $P_{\rm S}(t)$ is equal to the density of $A$ particles
on the lattice. Indeed, the asymptotic scaling properties of $P_{\rm S}(t)$ and $\rho(t)$
are the same.

To analyze the long time limit of $P_{\rm S}(t)$ 
in the high-density phase, where $1/2 < p < 1$,
one needs to use an asymptotic expansion of the hypergeometric function,
given by Eq. (\ref{eq:2F1}),
in which two of its indices, $a$ and $c$, tend to infinity simultaneously.
Such an expansion is given by
equation 3.13 in Ref. \cite{Paris2013} 

\begin{eqnarray}
\, _2 F_1
\left(\left. \begin{array}{c}
a+\epsilon t,b \\
c+t 
\end{array}
\right|
z \right)
&\simeq&
\frac{ \Gamma(c+t) \Gamma[1-c+a+(\epsilon-1)t] }{\Gamma(a+\epsilon t)}
\times
\nonumber \\
& &
\frac{ \epsilon^{a - 1/2 + \epsilon t} (\epsilon-1)^{c-a-1/2+(1-\epsilon)t}}{\sqrt{2 \pi t}(1-\epsilon z)^b},
\label{eq:Paris}
\end{eqnarray}

\noindent
where $\Gamma(x)$ is the Gamma function
\cite{Olver2010} and $t \gg 1$.
This expansion is valid for any $\epsilon > 0$, in the range of $0 < z < 1/\epsilon$.
Comparing the hypergeometric functions in Eqs. (\ref{eq:PSt_hg}) and (\ref{eq:Paris}),
we find that this expansion can be applied in the case of $1/2 < p < 1$,
where it yields 

\begin{equation}
\, _2 F_1
\left(\left. \begin{array}{c}
2,2t+1 \\
t+2 
\end{array}
\right|
1-p \right)
\simeq
\frac{1}{(2p-1)^2} +  \mathcal{O} \left( \frac{1}{t} \right).
\label{eq:2F1b}
\end{equation}

\noindent
Inserting the expansion of the function $\, _2 F_1(\ )$ 
from Eq. (\ref{eq:2F1b}) into Eq. (\ref{eq:PSt_hg})
and expanding the binomial coefficient in powers of $1/t$, 
we obtain

\begin{equation}
P_{\rm S}(t) \simeq \frac{2p-1}{p}
+ \frac{ 1-p }{\sqrt{\pi} (2p-1)^2}
\frac{ [4p(1-p)]^t }{t^{3/2}}.
\end{equation}

\noindent
Clearly, in the infinite time limit, $P_{\rm S}(t)$ converges
to an asymptotic value of $(2p-1)/p$.
The convergence follows the form

\begin{equation}
P_{\rm S}(t) 
\simeq
\frac{2p-1}{p}
+
\frac{ 1-p }{4 \sqrt{\pi} (p-1/2)^2}
\frac{ e^{-t/\tau(p)} }{t^{3/2}},
\label{eq:PSthdpltl1}
\end{equation}

\noindent
where the relaxation time is given by

\begin{equation}
\tau(p)= - \frac{1}{ \ln[4p(1-p)] }.
\label{eq:taup1}
\end{equation}

\noindent
Expanding $\tau(p)$ in powers of $p-1/2$, we obtain

\begin{equation}
\tau(p) = \frac{1}{4 (p-1/2)^2} + {\mathcal O}(1).
\label{eq:taup2}
\end{equation}

\noindent
This implies that in the limit of $p \rightarrow 1/2^{+}$
the relaxation time $\tau(p)$ diverges quadratically in
$1/(p-1/2)$.
This is similar to the critical slowing down observed in 
continuous phase transitions in equilibrium.

Applying the same asymptotic expansion to Eq. (\ref{eq:PSt_hg2}),
we find that for
$0 < p < 1/2$,
in the long time limit,
the stopping probability satisfies

\begin{equation}
P_{\rm S}(t) \simeq
\frac{ 1-p }{\sqrt{\pi} (1-2p)^2}
\frac{ [4p(1-p)]^t }{t^{3/2}},
\label{eq:PSt_ltl}
\end{equation}

\noindent
which can also be expressed in the form

\begin{equation}
P_{\rm S}(t)  
\simeq
\frac{ 1-p }{4 \sqrt{\pi} (1/2-p)^2}
\frac{ e^{-t/\tau(p)} }{t^{3/2}},
\label{eq:PSt_ltl2}
\end{equation}

\noindent
where $\tau(p)$ is given by Eq. (\ref{eq:taup1}).
Thus, using Eq. (\ref{eq:taup2}) one concludes that
in the critical regime on both sides of the transition point 
$p_c=1/2$,
the relaxation time
can be expressed in the form

\begin{equation}
\tau(p) \propto \left| p - p_c \right|^{- \gamma},
\end{equation}

\noindent
where the critical exponent is $\gamma = 2$.
This is in agreement with the scaling of the diverging relaxation time 
shown in equation (6) of Ref. \cite{Jha2025}.

Taking the limit of $p \rightarrow 1/2^{+}$ in Eq. (\ref{eq:PSthdpltl1})
and the limit of $p \rightarrow 1/2^{-}$ in Eq. (\ref{eq:PSt_ltl2}),
it is found that in both cases the stopping probability
exhibits a power-law decay of the form
$P_{\rm S}(t) \sim t^{-3/2}$ 
in the long-time limit.
At the transition point $p=1/2$ itself, the stopping probability
also exhibits a power-law decay in the long-time limit, which is given by 
$P_{\rm S}(t) \sim t^{-1/2}$ 
[see Eq. (\ref{eq:PStaprx})].
The smaller value of the power-law exponent implies that at the
transition point itself the stopping probability decays much more
slowly than at any point in its close vicinity on either side.

Another useful way to explore the probability $P_{\rm S}(t)$ is by
expanding the term $(1-p)^{\ell}$ on the right-hand side 
of Eq. (\ref{eq:PSt2}) in powers of $p$
and rearranging terms, we obtain

\begin{equation}
P_{\rm S}(t) = 2 \binom{2t-1}{t-1} \sum_{k=0}^{t-1} (-1)^k \frac{t-k}{(t+k)(t+k+1)} \binom{t}{k} p^{t+k}.
\label{eq:PStbsrbp}
\end{equation}

\noindent
Expressing the coefficient in front of the summation in Eq. (\ref{eq:PStbsrbp})
in terms of the Catalan number, 
given by Eq. (\ref{eq:Catalan}),
we obtain

\begin{equation}
P_{\rm S}(t) =    
C_t \sum_{k=0}^{t-1} 
(-1)^k \frac{(t-k)(t+1)}{(t+k)(t+k+1)} 
\binom{t}{k} p^{t+k}.
\label{eq:PStCt}
\end{equation}

\noindent
Writing the sum on the right-hand side 
of Eq. (\ref{eq:PStCt})
explicitly,
we obtain

\begin{equation}
P_{\rm S}(t) = C_t \left[ p^t - \frac{t(t-1)}{t+2} p^{t+1} + \frac{ t(t-1)(t-2) }{ 2 (t+2)(t+3) } p^{t+2}
+ \dots + (-1)^{t-1} \frac{  t+1  }{ 2(2t-1) } p^{2t-1} \right].
\label{eq:PStCt2}
\end{equation}

\noindent
In the low-density limit of $p \rightarrow 0$, the relaxation process is dominated by the leading term
on the right-hand side of Eq. (\ref{eq:PStCt2}), 
namely

\begin{equation}
P_{\rm S}(t) \simeq C_t  p^t.
\end{equation}

\noindent
In this limit the rate of stopping events follows an exponential decay.

\begin{figure}
\includegraphics[width=8.1cm]{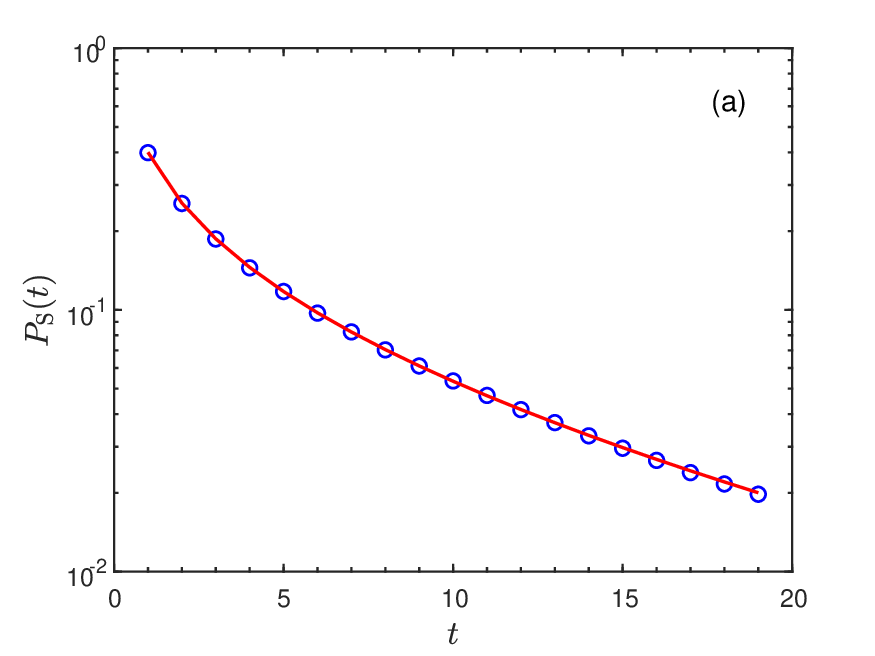} 
\includegraphics[width=8.1cm]{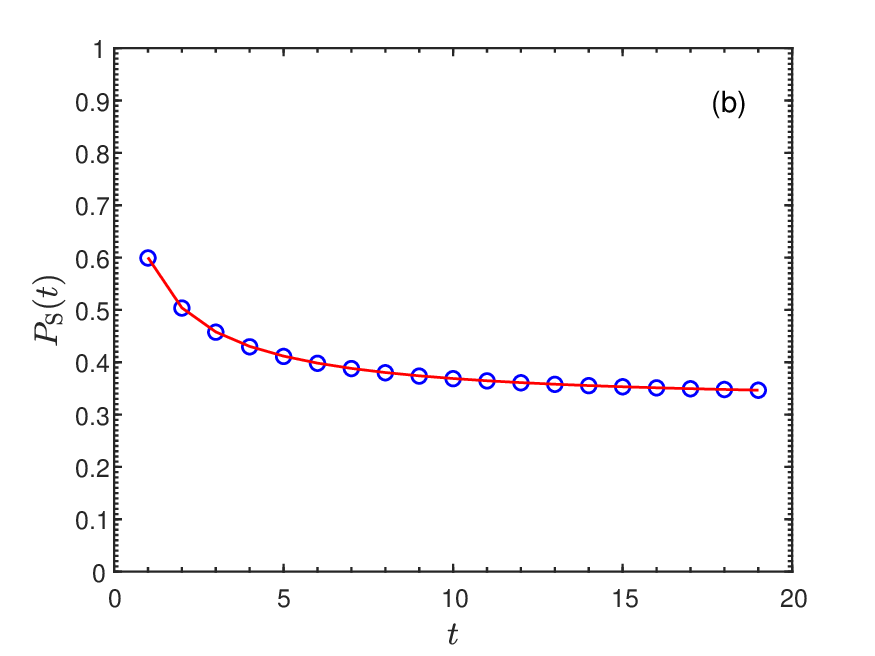} 
\caption{
(a) Analytical results (solid line) for the probability
$P_{\rm S}(t)$,
obtained from Eq. (\ref{eq:PSt_hg2}),
that a randomly selected car will be stopped at time $t$,
for a random initial state of density $p=0.4$.
In the low-density phase $P_{\rm S}(t)$
follows an exponentially truncated power-law distribution
and decays to zero in the long-time limit.
(b) Analytical results (solid line) for the probability
$P_{\rm S}(t)$
obtained from Eq. (\ref{eq:PSt_hg}),
that a randomly selected car will be stopped at time $t$,
for a random initial state of density $p=0.6$.
In the high-density phase $P_{\rm S}(t)$
follows an exponentially truncated power-law distribution
and converges towards a non-zero stationary value in the long-time limit. 
The analytical results are in very good agreement with the results obtained from
computer simulations (circles) on a lattice of size $L=10,000$.
}
\label{fig:7}
\end{figure}

In Fig. \ref{fig:7}(a) we present
analytical results (solid line) for the probability
$P_{\rm S}(t)$,
obtained from Eq. (\ref{eq:PSt_hg2}),
that a randomly selected car 
will be stopped at time $t$,
for a random initial state of density $p=0.4$
(low-density phase).
In Fig. \ref{fig:7}(b) we present
analytical results (solid line) for the probability
$P_{\rm S}(t)$
obtained from Eq. (\ref{eq:PSt_hg}),
that a randomly selected car will be stopped at time $t$,
for a random initial state of density $p=0.6$
(high-density phase).
The analytical results are in very good agreement with the results obtained from
computer simulations (circles) on a lattice of size $L=10,000$.
The simulation results are averaged over 50 initial configurations.

\begin{figure}
\includegraphics[width=9.0cm]{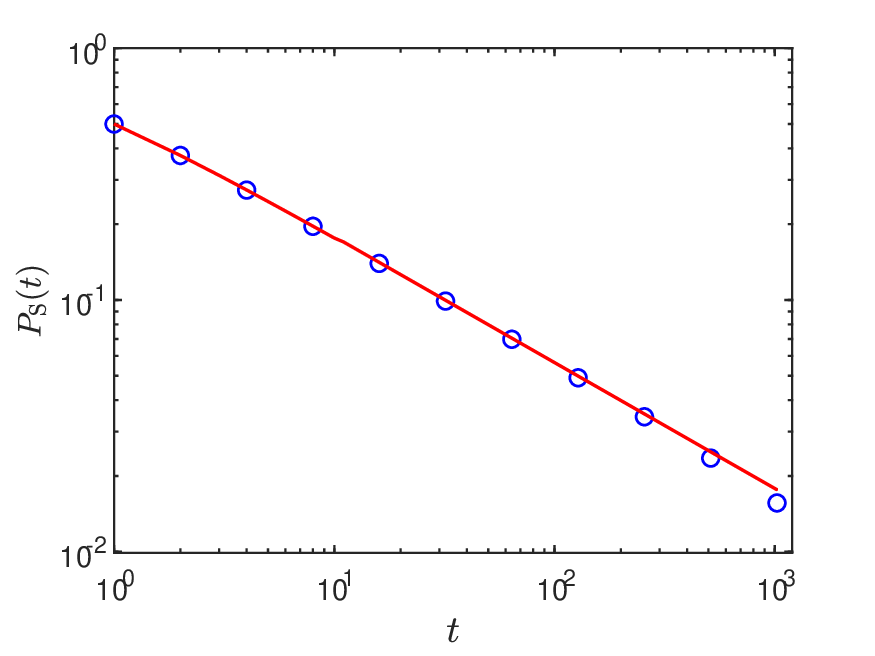} 
\caption{
Analytical results (solid line) for the probability
$P_{\rm S}(t)$ 
obtained from Eq. (\ref{eq:PSt12}),
that a randomly selected car will be stopped at time $t$
for a random initial state,
in the special case in which the car density is $p=1/2$
(at the transition point).
In this case $P_{\rm S}(t)$ exhibits a power-law decay with
no exponential truncation.
The analytical results are in very good agreement with the results obtained from
computer simulations (circles) on a lattice of size $L=10,000$.
}
\label{fig:8}
\end{figure}

In Fig. \ref{fig:8}  we present
analytical results (solid line) for the probability
$P_{\rm S}(t)$ 
obtained from Eq. (\ref{eq:PSt12}),
that a randomly selected car will be stopped at time $t$
for a random initial state of density $p=1/2$
(at the transition point).
The analytical results are in very good agreement with the results obtained from
computer simulations (circles) on a lattice of size $L=10,000$.
The simulation results are averaged over 1,000 initial configurations.

\section{The distribution of last-stopping times}

For car densities in the range of $0 < p < 1/2$ each car 
may be stopped for only a finite number of times.
The last time at which a given car is stopped is referred to as 
its last-stopping time.
Below we derive a closed-form expression for the
probability $P(T_{\rm LS}=t)$ of LS times. 
This is the probability that a
randomly selected car will be stopped at time $t$ and will then flow freely with no obstruction
at all the subsequent time steps.
The probability that a car will stop at time $t$ is determined by the initial configuration
of occupied and empty cells within an interval of $2t-1$ cells in front of that car. 
The car will be stopped at time $t$ if this interval contains at least $t$ cars at $t=0$,
whose initial arrangement satisfies the three conditions listed in Sec. IV.
We label the first $t$ cars ahead of the randomly selected car,
from left to right by $j=1,2,\dots,t$.
In case that at a later time $t' > t$ the car $j=t$ will stop, it will cause the randomly selected
car to stop again at some later time. 
However, if the car $j=t$ is one of the cars that
flow freely from the start, it implies that the randomly selected car will stop for the
last time at time $t$ and will flow freely afterwards.
The probability $P_{\rm NS}$ that the car $j=t$ will flow freely from the start 
is given by Eq. (\ref{eq:PNS1}).
Therefore, the probability that a randomly selected car will stop for the
last time at time $t$ is given by

\begin{equation}
P(T_{\rm LS}=t) =     \frac{  P_{\rm NS}   }{ 1 - P_{\rm NS} } P_{\rm S}(t),
\label{eq:PTLSt}
\end{equation}

\noindent
where the division by $1 - P_{\rm NS}$ accounts for the condition that
in order to be able to stop for the last time, a car must stop at least once 
and the multiplication by $P_{\rm NS}$ accounts for the probability that
the car in front of the chain of cars moves freely from the start.
Inserting $P_{\rm NS}$ from Eq. (\ref{eq:PNS1}) and $P_{\rm S}(t)$ from Eq. (\ref{eq:PSt2})
into Eq. (\ref{eq:PTLSt}), 
we obtain

\begin{equation}
P(T_{\rm LS}=t) = \frac{(1-2p) p^{t-1}}{t}
\sum_{\ell=0}^{t-1} (t-\ell) \binom{t-1+\ell}{t-1} (1-p)^{\ell},
\label{eq:PTLSt2}
\end{equation}

\noindent
where $t \ge 1$.
Expressing the right-hand side of Eq. (\ref{eq:PTLSt2})
in terms of a hypergeometric function, we obtain

\begin{equation}
P(T_{\rm LS}=t) = 
\frac{1-2p}{p^2}
\binom{2t}{t-1} 
\frac{ [p(1-p)]^{t+1} }{t}
\, _2 F_1
\left(\left. \begin{array}{c}
2,2 t+1 \\
t + 2
\end{array}
\right|
p  \right).
\label{eq:PLSt_hg2}
\end{equation}

To obtain an asymptotic expression for $P(T_{\rm LS}=t)$ that is valid in the
long-time limit, we insert $P_{\rm S}(t)$ from Eq. (\ref{eq:PSt_ltl}) into Eq. (\ref{eq:PTLSt})
and obtain

\begin{equation}
P(T_{\rm LS}=t) \simeq 
\frac{ 1-p }{\sqrt{\pi} p (1-2p)}
\frac{ [4p(1-p)]^t }{t^{3/2}},
\label{eq:PTLSt3}
\end{equation}
 
\noindent
which implies that in the long-time limit

\begin{equation}
P(T_{\rm LS}=t) 
\sim 
\frac{ e^{-t/\tau(p)} }{t^{3/2}},
\label{eq:PTLSt4}
\end{equation}

\noindent
where $\tau(p)$ is given by Eq. (\ref{eq:taup1}).

\begin{figure}
\includegraphics[width=9.0cm]{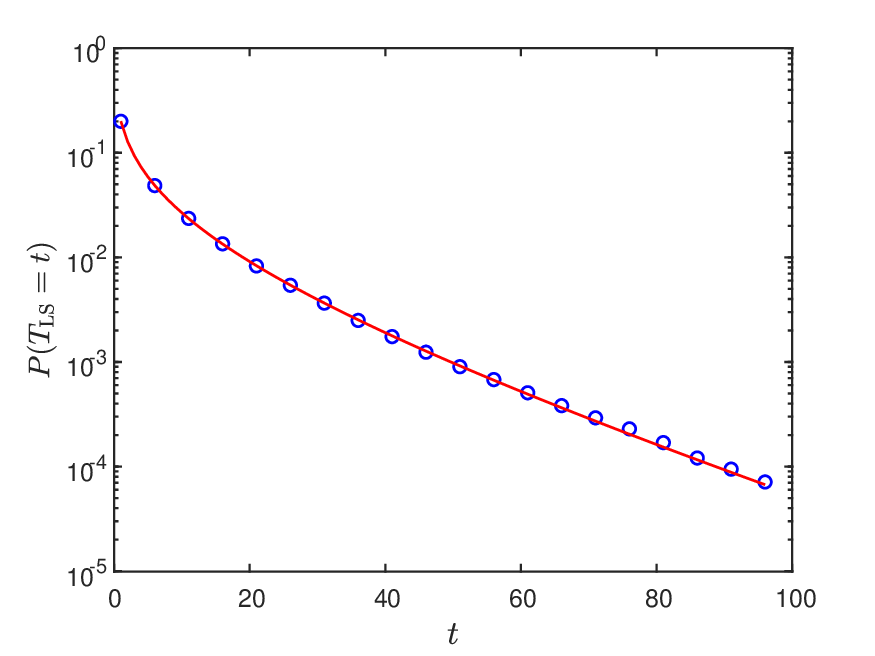} 
\caption{
Analytical results (solid line) for the distribution 
$P(T_{\rm LS}=t)$
of LS times 
in the low-density phase
of the DSTF model 
for a random initial state of density $p=0.4$.
The analytical results,
obtained from Eq. (\ref{eq:PLSt_hg2}),
are in very good agreement with the results obtained from
computer simulations (circles) on a lattice of size $L=10,000$.
}
\label{fig:9}
\end{figure}

In Fig. \ref{fig:9} we present 
analytical results (solid line) for the distribution 
$P(T_{\rm LS}=t)$
of LS times
in the low-density phase of the DSTF model
for a random initial state of density $p=0.4$.
The analytical results,
obtained from Eq. (\ref{eq:PLSt_hg2}),
are in very good agreement with the results obtained from
computer simulations (circles) on a lattice of size $L=10,000$.
The simulation results are averaged over 500 initial configurations.

The generating function of $P(T_{\rm LS}=t)$ is given by

\begin{equation}
R(z) = \sum_{t=1}^{\infty} z^t P(T_{\rm LS}=t).
\label{eq:GenFun}
\end{equation}
 
\noindent 
In Appendix B we show that

\begin{equation}
R(z) =  \frac{1-2p}{p}   
\frac{ 1 - \sqrt{ 1 - 4 zp(1-p) } }{ 1 - 2p + \sqrt{ 1 - 4 zp(1-p) } }. 
\label{eq:Rz}
\end{equation}

\noindent
In case that $0 < p < 1/2$ the radius of convergence is larger than $1$.
Thus, for $0 < p < 1/2$ one can safely insert
$z=1$ in Eq. (\ref{eq:Rz}) 
as it lies inside the radius of convergence.
This yields

\begin{equation}
R(1) = \sum_{t=1}^{\infty} P(T_{\rm LS}=t) = 1,
\label{eq:R1}
\end{equation} 

\noindent
which confirms the normalization of $P(T_{\rm LS}=t)$.
It is also observed that $R(1)$ diverges 
for $p \ge 1/2$.
 
Using the generating function $R(z)$, given by Eq. (\ref{eq:Rz}),
one can calculate the first two moments 
of $P(T_{\rm LS}=t)$.
It is found that the mean LS time is given by

\begin{equation}
{\mathbb E}[T_{\rm LS}] = 
\frac{ d R(z) }{d z} \bigg\vert_{z=1} =
\frac{ (1 - p)^2 }{ (1-2p)^2 }.
\label{eq:ETLSTLSi}
\end{equation}

\noindent
In the low-density limit of $p \rightarrow 0$ the mean LS time satisfies
${\mathbb E}[T_{\rm LS}] \rightarrow 1$,
which implies that in this limit most of the LS 
events take place at time $t=1$.
Note that the
mean LS time increases monotonically as $p$ is increased
and diverges at $p \rightarrow 1/2^{-}$.

\begin{figure}
\includegraphics[width=9.0cm]{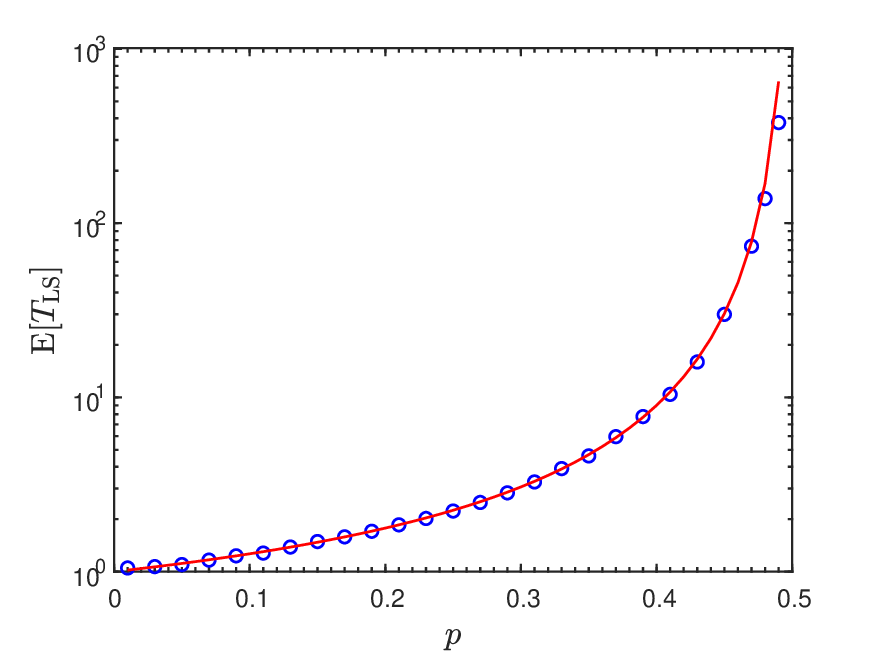} 
\caption{
Analytical results (solid line) for the expectation value
${\mathbb E}[T_{\rm LS}]$
of LS times in the low-density phase of the DSTF model, 
as a function of the car density $0 < p < 1/2$.
The analytical results, obtained from Eq. (\ref{eq:ETLSTLSi}),
are in very good agreement with the results obtained
from computer simulations (circles) on a lattice of size $L=10,000$.
As can be seen,
the mean LS time increases monotonically as a function of $p$
and diverges at $p \rightarrow 1/2^{-}$.
}
\label{fig:10}
\end{figure}

In Fig. \ref{fig:10} we present
analytical results (solid line) for the expectation value
${\mathbb E}[T_{\rm LS}]$
of LS times in the low-density phase of the DSTF model, 
as a function of the car density $0 < p < 1/2$.
As can be seen, the
mean LS time increases monotonically as a function of $p$
and diverges at $p \rightarrow 1/2^{-}$.
The analytical results, obtained from Eq. (\ref{eq:ETLSTLSi}),
are in very good agreement with the results obtained
from computer simulations (circles) on a lattice of size $L=10,000$.
The simulation results were averaged over 100 initial configurations
for each value of $p$.

The second moment 
of $P(T_{\rm LS}=t)$
is given by

\begin{equation}
{\mathbb E}[T_{\rm LS}^2] =
\frac{d}{dz} \left[ z \frac{ d R(z) }{d z} \right] \bigg\vert_{z=1} =
\frac{ (1-p)^2 }{ (1-2p)^4 }.
\label{eq:ETLS2pg12}
\end{equation}

\noindent
Thus, the variance is

\begin{equation}
{\rm Var}(T_{\rm LS}) =
\frac{ p (1-p)^2 (2-p) }{ (1-2p)^4  }.
\label{eq:VTLSTLSi}
\end{equation}

\noindent
In the low-density limit of $p \rightarrow 0$ the variance vanishes.
It increases monotonically as $p$ is increased and diverges at $p \rightarrow 1/2^{-}$.

\section{The joint distribution of last-stopping times and of the number of stopping events}

We begin this section with the observation that the summation variable
$\ell$ on the right-hand side of Eq. (\ref{eq:PTLSt2})
represents the number of empty cells between the randomly 
selected car and the car labeled by $j=t$.
In fact, the number of times in which the randomly selected car will stop
during the first $t$ time steps is given by $n = t-\ell$.
Using this observation, we can identify the joint distribution
$P(T_{\rm LS}=t,N_{\rm S}=n | N_{\rm S} \ge 1)$,
which is the probability that a randomly selected car will stop for the last time at time $t$
and that the total number of times that it will stop up to time $t$  
will be $n$, under the condition that $n \ge 1$.
This probability is given by

\begin{equation}
P(T_{\rm LS}=t,N_{\rm S}=n | N_{\rm S} \ge 1) = \frac{ n (1-2p) p^{t-1}}{t} \binom{2t-1-n}{t-1} (1-p)^{t-n},
\label{eq:PTLStNSn}
\end{equation}

\noindent
where $t \ge n$.
From this joint distribution
$P(T_{\rm LS}=t,N_{\rm S}=n | N_{\rm S} \ge 1)$
one can obtain the marginal distribution
$P(N_{\rm S}=n | N_{\rm S} \ge 1)$
of the number of stopping events experienced by a randomly
selected car before the system converges to an FFP state.
Tracing over the LS times, we obtain
the marginal distribution

\begin{equation}
P(N_{\rm S}=n | N_{\rm S} \ge 1) = \sum_{t=n}^{\infty} P(T_{\rm LS}=t,N_{\rm S} = n | N_{\rm S} \ge 1).
\label{eq:PNSn}
\end{equation}

\noindent
Inserting $P(T_{\rm LS}=t,N_{\rm S}=n | N_{\rm S} \ge 1)$ 
from Eq. (\ref{eq:PTLStNSn}) into Eq. (\ref{eq:PNSn}),
we obtain

\begin{equation}
P(N_{\rm S}=n | N_{\rm S} \ge 1) = \frac{ n(1-2p) }{p (1-p)^{n} } 
\sum_{t=n}^{\infty} \frac{p^t}{t} \binom{2t-1-n}{t-1} (1-p)^t.
\label{eq:PNSn2}
\end{equation}

\noindent
Eq. (\ref{eq:PNSn2}) can also be expressed in terms of a hypergeometric function,
in the form

\begin{equation}
P(N_{\rm S}=n | N_{\rm S} \ge 1) =
\frac{1-2p}{p}
p^n
\, _2 F_1
\left(\left. \begin{array}{c}
\frac{n}{2},\frac{n+1}{2} \\
n+1
\end{array}
\right|
4 p (1 - p) \right).
\end{equation}

\noindent
Using the identity 
(equation 15.4.17 in Ref. \cite{Olver2010})

\begin{equation}
\, _2 F_1
\left(\left. \begin{array}{c}
a,a+\frac{1}{2} \\
2a+1
\end{array}
\right|
z \right)
=
\left( \frac{2}{1+\sqrt{1-z}} \right)^{2a},
\end{equation}

\noindent
and keeping in mind that we focus here on the case of 
the low-density phase, where $0 < p < 1/2$,
we obtain

\begin{equation}
P(N_{\rm S}=n|N_{\rm S} \ge 1) = 
\frac{1-2p}{1-p} 
\left( \frac{p}{1-p} \right)^{n-1}.
\label{eq:PNSn5}
\end{equation}

\noindent
Thus, the distribution of the number of stopping events experienced by
a randomly selected car before the system converges to an FFP state
is simply a geometric distribution.

The mean of the conditional distribution 
$P(N_{\rm S}=n|N_{\rm S} \ge 1)$
is given by

\begin{equation}
{\mathbb E}[N_{\rm S} | N_{\rm S} \ge 1] = \frac{1-p}{1-2p},
\label{eq:ENSNSge1}
\end{equation}

\noindent
which converges to $1$
in the low-density limit of $p \rightarrow 0$ and
diverges in the limit of $p \rightarrow 1/2^{-}$.
The variance is given by

\begin{equation}
{\rm Var}(N_{\rm S} | N_{\rm S} \ge 1) = \frac{p(1-p)}{(1-2p)^2},
\label{eq:VarNSNSge1}
\end{equation}

\noindent
which vanishes in the limit of $p \rightarrow 0$ and
diverges in the limit of $p \rightarrow 1/2^{-}$.

The probability 
that a randomly selected car is never stopped satisfies

\begin{equation}
P(N_{\rm S}=0) = P_{\rm NS} = \frac{1-2p}{1-p},
\label{eq:PNS0}
\end{equation}

\noindent
where $P_{\rm NS}$ is given by Eq. (\ref{eq:PNS1}).
Combining Eqs. (\ref{eq:PNSn5}) and (\ref{eq:PNS0})
and using the law of total probability,
we obtain the unconditional distribution of the number
of stopping events, which is given by

\begin{equation}
P(N_{\rm S}=n) = 
\frac{1-2p}{1-p} 
\left( \frac{p}{1-p} \right)^{n},
\label{eq:PNSn6}
\end{equation}

\noindent
where $n \ge 0$.
The mean number of stopping events is given by

\begin{equation}
{\mathbb E}[N_{\rm S}] = \frac{p}{1-2p},
\end{equation}

\noindent
which 
vanishes in the low-density limit of $p \rightarrow 0$ and
diverges in the limit of $p \rightarrow 1/2^{-}$.
The variance of the (unconditional) distribution $P(N_{\rm S}=n)$ is given by

\begin{equation}
{\rm Var}(N_{\rm S}) = \frac{p(1-p)}{(1-2p)^2},
\end{equation}

\noindent
which vanishes in the low-density limit of $p \rightarrow 0$ and
diverges in the limit of $p \rightarrow 1/2^{-}$.

Another interesting quantity is the
conditional distribution of LS times under the condition that
the number of stopping events is $n$.
Using Bayes' theorem, it is given by

\begin{equation}
P(T_{\rm LS}=t | N_{\rm S}=n) =
\frac{ P(T_{\rm LS}=t,N_{\rm S}=n | N_{\rm S} \ge 1) }{ P(N_{\rm S}=n | N_{\rm S} \ge 1) }.
\label{eq:PTLStNSn2}
\end{equation}

\noindent
Inserting $P(T_{\rm LS}=t,N_{\rm S}=n | N_{\rm S} \ge 1)$ from Eq. (\ref{eq:PTLStNSn})
and $P(N_{\rm S}=n | N_{\rm S} \ge 1)$ from Eq. (\ref{eq:PNSn5}) into the right-hand side
of Eq. (\ref{eq:PTLStNSn2}) and rearranging terms, we obtain

\begin{equation}
P(T_{\rm LS}=t | N_{\rm S}=n) =
\frac{n}{t} \binom{2t-1-n}{t-1} p^{t-n} (1-p)^t,
\label{eq:PTLStNSn3}
\end{equation}

\noindent
for $n \ge 1$ and $t \ge n$.

\begin{figure}
\includegraphics[width=9.0cm]{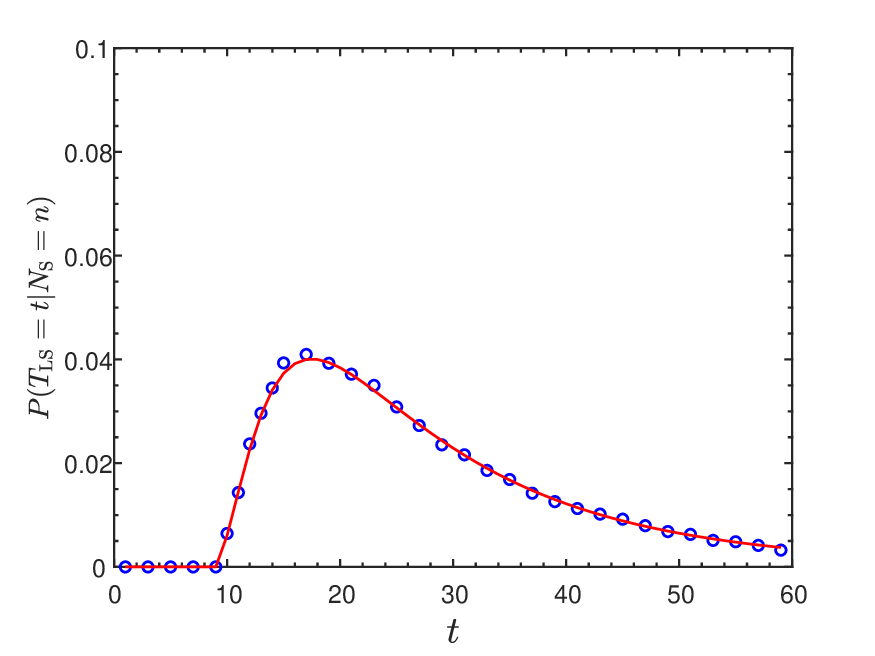} 
\caption{
Analytical results (solid line) for the conditional distribution
$P(T_{\rm LS}=t | N_{\rm S}=n)$
of LS times 
in the DSTF model with car density of $p=0.4$,
under the condition that the number of stopping events is $n=10$.
The analytical results, obtained from Eq. (\ref{eq:PTLStNSn3}),
are in very good agreement with the results obtained
from computer simulations (circles) on a lattice of size $L=10,000$.
}
\label{fig:11}
\end{figure}

In Fig. \ref{fig:11} we present
analytical results (solid line) for the conditional distribution
$P(T_{\rm LS}=t | N_{\rm S}=n)$
of LS times 
in the DSTF model with car density of $p=0.4$,
under the condition that the number of stopping events is $n$.
The analytical results, obtained from Eq. (\ref{eq:PTLStNSn3}),
are in very good agreement with the results obtained
from computer simulations (circles) on a lattice of size $L=10,000$.
The simulation results were averaged over 2,000 initial configurations.

Using the approach presented in Appendix B, 
it is found that the
generating function of the conditional distribution
$P(T_{\rm LS}=t | N_{\rm S}=n)$
is given by

\begin{equation}
U_{n}(z) = \left[ \frac{1-\sqrt{1-4 (1-p) p z}}{2p} \right]^n.
\end{equation}

\noindent
The first moment (mean) of this conditional distribution 
is given by

\begin{equation}
{\mathbb E}[T_{\rm LS} | N_{\rm S}=n] =
\frac{d}{dz} U_{n}(z) \bigg|_{z=1}.
\end{equation}

\noindent
Carrying out the differentiation and inserting $z=1$,
it is found that

\begin{equation}
{\mathbb E}[T_{\rm LS} | N_{\rm S}=n] =
\frac{1-p}{1-2p} n,
\label{eq:ETLSNSn}
\end{equation}

\noindent
where $n \ge 1$.
This implies that the expectation value of the LS time
increases linearly with the number of stopping events.
The pre-factor is a monotonically increasing function of $p$,
which diverges in the limit of 
$p \rightarrow 1/2^{-}$.

\begin{figure}
\includegraphics[width=9.0cm]{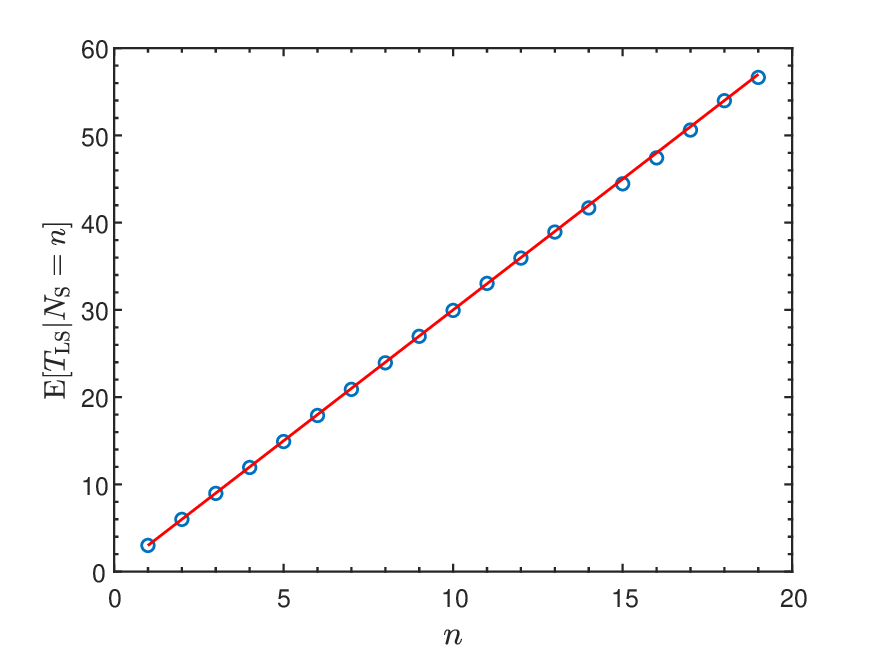} 
\caption{
Analytical results (solid line) for the conditional expectation value
${\mathbb E}[T_{\rm LS}  | N_{\rm S}=n]$
of LS times 
in the DSTF model with car density of $p=0.4$,
under the condition that the number of stopping events is $n$,
as a function of $n$.
The analytical results, obtained from Eq. (\ref{eq:ETLSNSn}),
are in very good agreement with the results obtained
from computer simulations (circles) on a lattice of size $L=10,000$.
}
\label{fig:12}
\end{figure}

In Fig. \ref{fig:12} we present
analytical results (solid line) for the conditional expectation value
${\mathbb E}[T_{\rm LS}  | N_{\rm S}=n]$
of LS times 
in the DSTF model with car density of $p=0.4$,
under the condition that the number of stopping events is $n$.
The analytical results, obtained from Eq. (\ref{eq:ETLSNSn}),
are in very good agreement with the results obtained
from computer simulations (circles) on a lattice of size $L=10,000$.
The simulation results were averaged over $10,000$ initial configurations.

The variance of
$P(T_{\rm LS}=t | N_{\rm S}=n)$
is given by

\begin{equation}
{\rm Var}(T_{\rm LS} | N_{\rm S}=n) =
\frac{d^2}{dz^2} U_{n}(z) \bigg|_{z=1} 
+
\frac{d}{dz} U_{n}(z) \bigg|_{z=1} 
-
\left[ \frac{d}{dz} U_{n}(z) \bigg|_{z=1} \right]^2.
\end{equation}

\noindent
Carrying out the differentiations and inserting $z=1$,
it is found that

\begin{equation}
{\rm Var}(T_{\rm LS} | N_{\rm S}=n) =
\frac{ p(1-p) }{ (1-2p)^3 } n,
\label{eq:VTLSNSn}
\end{equation}

\noindent
where $n \ge 1$.

Yet another interesting quantity is the opposite distribution, namely the
conditional distribution of the number of stopping events for
a given value of the LS time.
It is given by

\begin{equation}
P( N_{\rm S}=n | T_{\rm LS}=t ) =
\frac{ P(T_{\rm LS}=t,N_{\rm S}=n | N_{\rm S} \ge 1) }{ P(T_{\rm LS}=t) },
\label{eq:PNSnTLSt}
\end{equation}

\noindent
for $1 \le n \le t$.
Inserting $P(T_{\rm LS}=t,N_{\rm S}=n | N_{\rm S} \ge 1)$ from Eq. (\ref{eq:PTLStNSn})
and $P(T_{\rm LS}=t)$ from Eq. (\ref{eq:PLSt_hg2}) into the right-hand side
of Eq. (\ref{eq:PNSnTLSt}) and rearranging terms, we obtain

\begin{equation}
P(N_{\rm S}=n | T_{\rm LS}=t) =
\frac{  n   \binom{2t-1-n}{t-1} \left( \frac{1}{1-p} \right)^{n-1} }
{ \binom{2t-2}{t-1}  
\, _2F_1 \left( \left.
\begin{array}{c}
2, 1-t \\
2-2t
\end{array}
\right| \frac{1}{1-p} 
\right) 
}.
\label{eq:PNSnTLSt2}
\end{equation}

\begin{figure}
\includegraphics[width=9.0cm]{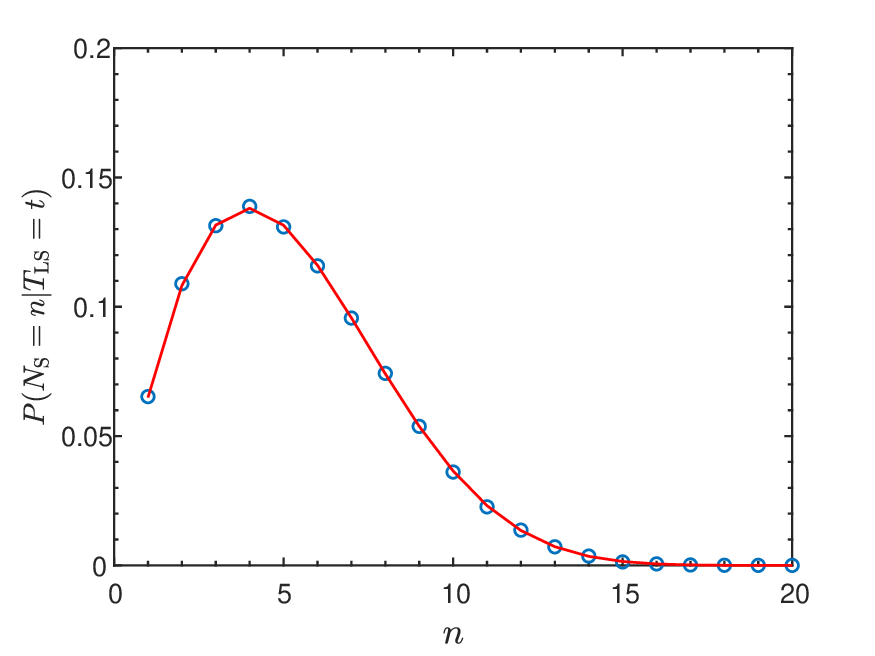} 
\caption{
Analytical results (solid line) for the conditional distribution
$P(N_{\rm S}=n | T_{\rm LS}=t)$
of the number of stopping events
in the DSTF model with car density of $p=0.4$,
under the condition that the LS time is $t=20$.
The analytical results, obtained from Eq. (\ref{eq:PNSnTLSt2}),
are in very good agreement with the results obtained
from computer simulations (circles) on a lattice of size $L=10,000$.
}
\label{fig:13}
\end{figure}

In Fig. \ref{fig:13} we present  
analytical results (solid line) for the conditional distribution
$P(N_{\rm S}=n | T_{\rm LS}=t)$
of the number of stopping events 
in the DSTF model with car density of $p=0.4$,
under the condition that the LS time is $t=20$.
The analytical results, obtained from Eq. (\ref{eq:PNSnTLSt2}),
are in very good agreement with the results obtained
from computer simulations (circles) on a lattice of size $L=10,000$. 
The simulation results were averaged over $10,000$ initial configurations.

Below we calculate the moment generating function of
$P(N_{\rm S}=n | T_{\rm LS}=t)$.
It is given by

\begin{equation}
M(s) = \sum_{n=1}^{\infty}
e^{s n} P(N_{\rm S}=n | T_{\rm LS}=t).
\label{eq:Ms}
\end{equation}

\noindent
Inserting
$P(N_{\rm S}=n | T_{\rm LS}=t)$
from Eq. (\ref{eq:PNSnTLSt2})
into Eq. (\ref{eq:Ms})
and carrying out the summation,
we obtain

\begin{equation}
M(s) = e^{s} 
\frac{ 
\, _2F_1 \left( \left.
\begin{array}{c}
2, 1-t \\
2-2t
\end{array}
\right| \frac{e^{s}}{1-p} 
\right) 
}
{
\, _2F_1 \left( \left.
\begin{array}{c}
2, 1-t \\
2-2t
\end{array}
\right| \frac{1}{1-p} 
\right)
}.
\label{eq:Ms2}
\end{equation}

\noindent
Taking the logarithm of $M(s)$, we obtain the cumulant-generating function,
which is given by

\begin{eqnarray}
K(s) = s 
&+& 
\ln \left[
\, _2F_1 \left( \left.
\begin{array}{c}
2, 1-t \\
2-2t
\end{array}
\right| \frac{e^{s}}{1-p} 
\right) 
\right]
\nonumber \\
&-&
\ln
\left[
\, _2F_1 \left( \left.
\begin{array}{c}
2, 1-t \\
2-2t
\end{array}
\right| \frac{1}{1-p} 
\right) 
\right].
\label{eq:Ks}
\end{eqnarray}

\noindent
The cumulants of
$P(N_{\rm S}=n | T_{\rm LS}=t)$
are given by

\begin{equation}
\kappa_i = \frac{ d^i K(s) }{d s^i} \bigg|_{s=0}.
\label{eq:kappak}
\end{equation}

\noindent
In particular, the mean is given by
${\mathbb E}[N_{\rm S} | T_{\rm LS}=t] = \kappa_1$
and the variance is given by
${\rm Var}(N_{\rm S} | T_{\rm LS}=t) = \kappa_2$.
Clearly, in the special case of $t=1$ one obtains
${\mathbb E}[N_{\rm S} | T_{\rm LS}=1] = 1$
and
${\rm Var}(N_{\rm S} | T_{\rm LS}=1) = 0$.

To calculate these cumulants, we apply repeated differentiation of $K(s)$,
which yields

\begin{equation}
\kappa_i = \delta_{i,1} +
\sum_{j=1}^i S(i,j) r^j \frac{d^j}{dr^j}
\ln \left[
\, _2F_1 \left( \left.
\begin{array}{c}
2, 1-t \\
2-2t
\end{array}
\right| r 
\right) 
\right],
\end{equation}

\noindent
where 

\begin{equation}
r = \frac{1}{1-p},
\end{equation}

\noindent
and the coefficients $S(i,j)$ are the Stirling numbers of the second kind
\cite{Olver2010}.

Using Eq. 15.5.2 in Ref. \cite{Olver2010}, we obtain
the identity

\begin{equation}
\frac{d^j}{dr^j}
\, _2F_1 \left( \left.
\begin{array}{c}
2, 1-t \\
2-2t
\end{array}
\right| r 
\right) 
=
\frac{(2)_j (1-t)_j}{(2-2t)_j}
\, _2F_1 \left( \left.
\begin{array}{c}
2+j, 1-t+j \\
2-2t+j
\end{array}
\right| r 
\right).
\label{eq:djF}
\end{equation}

\noindent
Using Eq. (\ref{eq:djF}) for $j=1$, 
we obtain

\begin{equation}
\frac{d}{dr}
\ln
\left[
\, _2F_1 \left( \left.
\begin{array}{c}
2, 1-t \\
2-2t
\end{array}
\right| r 
\right) 
\right]
=
\frac{
\, _2F_1 \left( \left.
\begin{array}{c}
3, 2-t \\
3-2t
\end{array}
\right| r 
\right) 
}
{
\, _2F_1 \left( \left.
\begin{array}{c}
2, 1-t \\
2-2t
\end{array}
\right| r 
\right) 
}.
\label{eq:drlnF}
\end{equation}

\noindent
Using Eq. (\ref{eq:djF}) for $j=1$ and $j=2$, 
we obtain

\begin{equation}
\frac{d^2}{dr^2}
\ln
\left[
\, _2F_1 \left( \left.
\begin{array}{c}
2, 1-t \\
2-2t
\end{array}
\right| r 
\right) 
\right]
=
\frac{3(t-2)}{2t-3}
\frac{
\, _2F_1 \left( \left.
\begin{array}{c}
4, 3-t \\
4-2t
\end{array}
\right| r 
\right) 
}
{
\, _2F_1 \left( \left.
\begin{array}{c}
2, 1-t \\
2-2t
\end{array}
\right| r 
\right) 
}
-
\left[
\frac{
\, _2F_1 \left( \left.
\begin{array}{c}
3, 2-t \\
3-2t
\end{array}
\right| r 
\right) 
}
{
\, _2F_1 \left( \left.
\begin{array}{c}
2, 1-t \\
2-2t
\end{array}
\right| r 
\right) 
}
\right]^2.
\label{eq:drlnF2}
\end{equation}

\noindent
Combining these results for $t \ge 2$, 
we obtain

\begin{equation}
{\mathbb E}[N_{\rm S}|T_{\rm LS}=t] = 
1 + \frac{1}{1-p} 
\frac{
\, _2F_1 \left( \left.
\begin{array}{c}
3, 2-t \\
3-2t
\end{array}
\right| \frac{1}{1-p} 
\right) 
}
{
\, _2F_1 \left( \left.
\begin{array}{c}
2, 1-t \\
2-2t
\end{array}
\right| \frac{1}{1-p} 
\right) 
},
\label{eq:PNSnTLStmean3}
\end{equation}

\noindent
and

\begin{eqnarray}
{\rm Var}(N_{\rm S}|T_{\rm LS}=t) &=& 
\frac{1}{1-p} 
\frac{
\, _2F_1 \left( \left.
\begin{array}{c}
3, 2-t \\
3-2t
\end{array}
\right| \frac{1}{1-p} 
\right) 
}
{
\, _2F_1 \left( \left.
\begin{array}{c}
2, 1-t \\
2-2t
\end{array}
\right| \frac{1}{1-p} 
\right) 
}
\nonumber \\
&+&
\frac{3}{(1-p)^2} 
\frac{t-2}{2t-3}
\frac{
\, _2F_1 \left( \left.
\begin{array}{c}
4, 3-t \\
4-2t
\end{array}
\right| \frac{1}{1-p} 
\right) 
}
{
\, _2F_1 \left( \left.
\begin{array}{c}
2, 1-t \\
2-2t
\end{array}
\right| \frac{1}{1-p} 
\right)
}
\nonumber \\
&-&
\frac{1}{(1-p)^2} 
\left[
\frac{
\, _2F_1 \left( \left.
\begin{array}{c}
3, 2-t \\
3-2t
\end{array}
\right| \frac{1}{1-p} 
\right) 
}
{
\, _2F_1 \left( \left.
\begin{array}{c}
2, 1-t \\
2-2t
\end{array}
\right| \frac{1}{1-p} 
\right)
}
\right]^2.
\label{eq:VarNSTLS}
\end{eqnarray}

In the long time limit, the expectation value of the conditional
distribution $P(N_{\rm S}=n | T_{\rm LS}=t)$
is given by

\begin{equation}
{\mathbb E}[N_{\rm S}|T_{\rm LS}=t] = 
\frac{3-2p}{1-2p} - 6 \frac{1-p}{(1-2p)^3} \frac{1}{t} + {\mathcal O} \left( \frac{1}{t^2} \right).
\label{eq:ENTt1}
\end{equation}

\noindent
This equation shows that the conditional expectation value 
increases as a function of the LS time toward a saturation value
of $(3-2p)/(1-2p)$. Thus, the expected number of stopping events
remains finite even when the last-stopping time diverges.
It implies that the trajectories of cars for which $T_{\rm LS}$ is very large
are dominated by a few periods of free flow between successive
stopping events.
Note that this is not an obvious outcome,
when keeping in mind that according to Eq. (\ref{eq:ETLSNSn}),
the conditional expectation value 
${\mathbb E}[T_{\rm LS} | N_{\rm S}=n]$
grows linearly with $n$.

Similarly, in the long time limit the variance is given by

\begin{equation}
{\rm Var}(N_{\rm S}|T_{\rm LS}=t) =
\frac{4(1-p)}{(1-2p)^2} - \frac{6(1-p)(5-4p)}{(1-2p)^4} \frac{1}{t}
+ {\mathcal O} \left( \frac{1}{t^2} \right).
\label{eq:VNSTt1}
\end{equation}

\begin{figure}
\includegraphics[width=9.0cm]{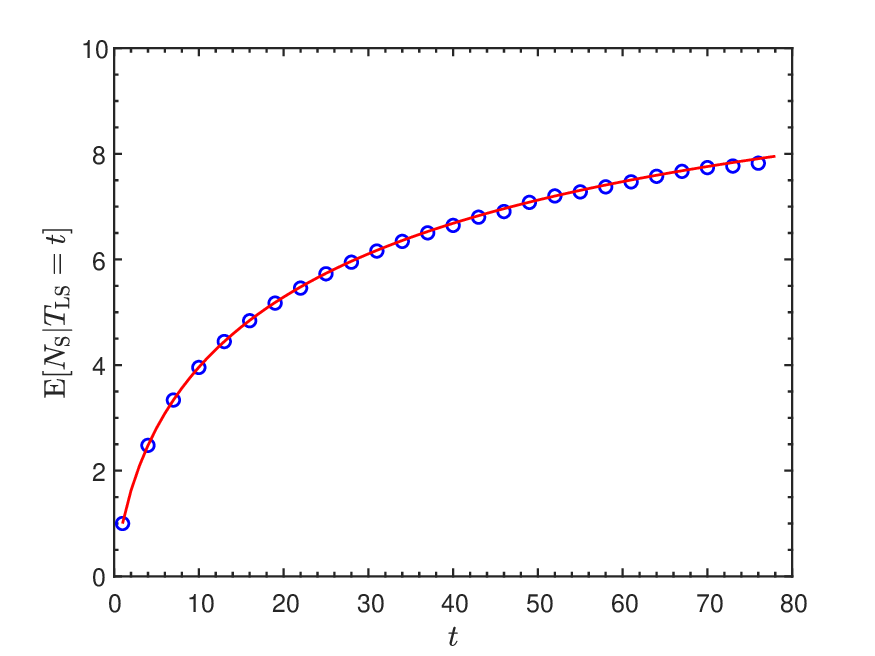} 
\caption{
Analytical results (solid line) for the conditional expectation value
${\mathbb E}[N_{\rm S}  | T_{\rm LS}=t]$
of the number of stopping events 
in the DSTF model with car density of $p=0.4$,
under the condition that the LS time is $t$.
The analytical results, obtained from Eq. (\ref{eq:PNSnTLStmean3}),
are in very good agreement with the results obtained
from computer simulations (circles) on a lattice of size $L=10,000$.
}
\label{fig:14}
\end{figure}

In Fig. \ref{fig:14} we present 
analytical results (solid line) for the conditional expectation value
${\mathbb E}[N_{\rm S}  | T_{\rm LS}=t]$
of the number of stopping events of a randomly selected car
in the DSTF model with car density of $p=0.4$,
under the condition that the LS time is $t$.
The analytical results, obtained from Eq. (\ref{eq:PNSnTLStmean3}),
are in very good agreement with the results obtained
from computer simulations (circles) on a lattice of size $L=10,000$.
The simulation results were averaged over $10,000$ initial configurations.

Using the Stirling formula,
for $n \gg 1$
Eq. (\ref{eq:PTLStNSn3}) 
can be approximated by

\begin{equation}
P(T_{\rm LS}=t | N_{\rm S}=n) = 
\left\{  
\begin{array}{ll}
(1-p)^n  & \ \ \ \ t=n \\
\frac{n}{\sqrt{2 \pi}} 
\frac{ (2t-n)^{2t-n-\frac{1}{2}} }{ (t-n)^{t-n+\frac{1}{2}} t^{ t+\frac{1}{2} } }
(1-p)^{t} p^{t-n} 
& \ \ \ \  t \ge n+1.
\end{array}
\right.
\end{equation}

\noindent
In the low-density limit, the conditional distribution is 
dominated by a fast-decaying exponential tail.
As the density $p$ is increased toward $p=1/2$, 
the peak of $P(T_{\rm LS}=t | N_{\rm S}=n)$ shifts to the right and
the exponential decay beyond the peak slows down, making 
the possibility of very long LS times more probable.

Below we consider the asymptotic form of the conditional distribution
$P(N_{\rm S}=n | T_{\rm LS}=t)$,
given by Eq. (\ref{eq:PNSnTLSt2}),
in the limit of $t \rightarrow \infty$.
In this limit, the hypergeometric function
on the right-hand side of Eq. (\ref{eq:PNSnTLSt2})
converges to

\begin{equation}
\, _2F_1 \left( \left.
\begin{array}{c}
2, 1-t \\
2-2t
\end{array}
\right| \frac{1}{1-p} 
\right) 
\xrightarrow[t \to \infty]{}
\frac{ 4 (1-p)^2 }{ (1-2p)^2 },
\end{equation}

\noindent
and

\begin{equation}
\frac{ \binom{2t-1-n}{t-1} }
{ \binom{2t-2}{t-1} }
\xrightarrow[t \to \infty]{}
\frac{1}{2^{n-1}}.
\end{equation}

\noindent
Inserting these results into the right-hand side of
Eq. (\ref{eq:PNSnTLSt2}),
it is found that in the asymptotic limit of $T_{\rm LS} \rightarrow \infty$,
the conditional distribution is reduced to

\begin{equation}
P(N_{\rm S}=n|T_{\rm LS} \rightarrow \infty) =
\frac{ (1-2p)^2 }{4 (1-p)^2} n \left[ \frac{1}{2 (1-p)} \right]^{ n-1 },
\label{eq:PNSTLSinf}
\end{equation}

\noindent
where $n \ge 1$,
which follows a negative binomial distribution
of the form
${\rm NB} \left[ 2, \frac{1-2p}{2(1-p)} \right]$.
The mean of this distribution is given by

\begin{equation}
{\mathbb E}[N_{\rm S}|T_{\rm LS} \rightarrow \infty] = 
\frac{3-2p}{1-2p},
\label{eq:ENTt2}
\end{equation}

\noindent
in agreement with the $t \rightarrow \infty$ limit 
of Eq. (\ref{eq:ENTt1}).
Similarly, the variance is given by

\begin{equation}
{\rm Var}(N_{\rm S}|T_{\rm LS} \rightarrow \infty) =
\frac{4(1-p)}{(1-2p)^2}, 
\label{eq:VNSTt2}
\end{equation}

\noindent
in agreement with the $t \rightarrow \infty$
limit of Eq. (\ref{eq:VNSTt1}).

We also calculate the correlation coefficient
$\rho(T_{\rm LS},N_{\rm S} | N_{\rm S} \ge 1)$.
To this end, we use the law of total covariance, which states that
if $X$, $Y$ and $Z$ are random variables on the same probability space and
the covariance of $X$ and $Y$ is finite, then
\cite{Ross2020}

\begin{equation}
{\rm Cov}(X,Y) = {\mathbb E}[{\rm Cov}(X,Y|Z)] + {\rm Cov}( {\mathbb E}[X|Z],{\mathbb E}[Y|Z] ).
\label{eq:cov1}
\end{equation}

\noindent
Inserting $X=T_{\rm LS}$, $Y=N_{\rm S}$ and $Z=N_{\rm S}$
in Eq. (\ref{eq:cov1}), we obtain

\begin{eqnarray}
{\rm Cov}(T_{\rm LS},N_{\rm S} | N_{\rm S} \ge 1) &=&
{\mathbb E}[ {\rm Cov}(T_{\rm LS},N_{\rm S} | N_{\rm S} =n) | N_{\rm S} \ge 1]
\nonumber \\
&+&
{\rm Cov} \left( {\mathbb E}[T_{\rm LS} | N_{\rm S}=n], 
{\mathbb E}[N_{\rm S} | N_{\rm S} = n]| N_{\rm S} \ge 1 \right).
\label{eq:cov1p}
\end{eqnarray}

\noindent
Inserting ${\mathbb E}[T_{\rm LS} | N_{\rm S}=n]$
from Eq. (\ref{eq:ETLSNSn}) into Eq. (\ref{eq:cov1p})
and using the fact that
${\rm Cov}(T_{\rm LS},N_{\rm S} | N_{\rm S}=n)=0$
and 
${\mathbb E}[N_{\rm S}|N_{\rm S}=n]=n$,
we obtain

\begin{equation}
{\rm Cov}(T_{\rm LS},N_{\rm S} | N_{\rm S} \ge 1) =
\frac{1-p}{1-2p} {\rm Var}(N_{\rm S} | N_{\rm S} \ge 1).
\label{eq:cov2}
\end{equation}

\noindent
Inserting ${\rm Var}(N_{\rm S} | N_{\rm S} \ge 1)$
from Eq. (\ref{eq:VarNSNSge1}) into Eq. (\ref{eq:cov2})
and carrying out the multiplication, we obtain

\begin{equation}
{\rm Cov}(T_{\rm LS},N_{\rm S} | N_{\rm S} \ge 1) =
\frac{p(1-p)^2}{(1-2p)^3}. 
\end{equation}

\noindent
To complete the calculation of the Pearson correlation coefficient

\begin{equation}
\rho(T_{\rm LS},N_{\rm S} | N_{\rm S} \ge 1) =
\frac{ {\rm Cov}(T_{\rm LS},N_{\rm S} | N_{\rm S} \ge 1) }
{ \sqrt{ {\rm Var}(T_{\rm LS} | N_{\rm S} \ge 1) } \sqrt{ {\rm Var}(N_{\rm S} | N_{\rm S} \ge 1) } },
\label{eq:rhoTN}
\end{equation}

\noindent
we also need closed-form expressions for 
${\rm Var}(N_{\rm S} | N_{\rm S} \ge 1)$,
which is given by Eq. (\ref{eq:VarNSNSge1}),
and for 
${\rm Var}(T_{\rm LS} | N_{\rm S} \ge 1)={\rm Var}(T_{\rm LS})$,
which is given by Eq. (\ref{eq:VTLSTLSi}).
Combining these results, we obtain

\begin{equation}
\rho(T_{\rm LS},N_{\rm S} | N_{\rm S} \ge 1) =
\sqrt{ \frac{1-p}{2-p} },
\label{eq:rhoTLSNS}
\end{equation}

\noindent
where $0 < p < 1/2$.
Thus, the correlation coefficient
takes its maximum value of 
$\rho(T_{\rm LS},N_{\rm S} | N_{\rm S} \ge 1) = 1/\sqrt{2}$
in the low-density limit of $p \rightarrow 0$.
It decreases monotonically as $p$ is increased,
towards the value of $1/\sqrt{3}$ at $p \rightarrow 1/2^{-}$.

\section{Discussion}

Mountain landscapes such as the one shown in Fig. \ref{fig:2} 
can be expressed as partial sums of random variables,
which are referred to as ladder processes
\cite{Feller1971a,Feller1971b}.
The record heights of $h(i)$, 
referred to as ascending ladder epochs,  
were studied in the context of fluctuation theory
\cite{Feller1971a,Feller1971b,Spitzer1970}.
An important insight obtained from these studies is that
ladder epochs are, in fact, regeneration epochs.
This means that at these points the function $h(i)$ is renewed in the
sense that the record that has been reached serves as a new starting
point for pursuing higher records.
As a result, the statistical properties of the trajectories between pairs
of successive records are all the same
from a statistical point of view.
This property explains the fact that the distribution $P(N_{\rm S}=n)$
of the number of stopping events,
given by Eq. (\ref{eq:PNSn6}),
follows a geometric distribution.
This is in spite of the fact that the joint distribution 
$P(T_{\rm LS}=t, N_{\rm S}=n)$,
from which it is derived,
is rather complex.

The evolution of the patterns of occupied and empty cells during the 
convergence to an FFP state was studied in Refs.
\cite{Fuks1997,Boccara1998,Fuks1999,Boccara2002,Fuks2023}.
Using finite state machines and pre-image analysis methods
they derived explicit solutions for the probabilities of finite block
patterns and their time evolution.
In particular, it was found that the decay of patterns that are not favored in
the FFP states resemble the decay of the stopping probability $P_{\rm S}(t)$,
given by Eq. (\ref{eq:PSt_hg}).

The results presented in this paper provide useful insight on the dynamics of
deterministic two-dimensional traffic-flow models such as the BML model
\cite{Biham1992}.
The BML model consists of horizontally oriented (H) cars that 
move to the right and vertically
oriented (V) cars that move downward, on a square lattice of 
size L with periodic boundary conditions. 
Starting from a random initial state of density $p$, which is 
equally divided between the H and V cars, the model exhibits
a first-order dynamical phase transition at a critical density $p_c$. 
For $p < p_c$ it evolves toward a free-flowing periodic state,
while for $p > p_c$, it evolves toward a fully jammed state 
or to an intermediate state of congested traffic.
The interactions between H cars in a single row or between V cars
in a single column are essentially described by the DSTF model.

The transient dynamics of the BML model in the low-density phase
was recently studied using a
configuration-space distance measure 
$D(t) = D_{\parallel}(t) + D_{\perp}(t)$
between the state of the system at time $t$
and the set of FFP states
\cite{Hertzberg2026}.
The $D_{\parallel}(t)$ term accounts for the interactions between homotypic pairs of H (or V) cars,
while $D_{\perp}(t)$ accounts for the interactions 
between heterotypic pairs of H and V-cars.
It was shown that in the FFP states $D(t)=0$, while in all the other states $D(t) > 0$.
It was also found that as the system evolves toward an FFP state,
there is a separation of time scales, where $D_{\parallel}(t)$ decays
very fast while $D_{\perp}(t)$ decays much more slowly.
In fact, the distance measure $D_{\parallel}(t)$,
which counts the number of pairs of homotypic cars that are adjacent to each other,
is equivalent to the
probability $P_{\rm S}(t)$ that a randomly selected car will be stopped 
at time $t$.

\section{Summary}

We presented analytical results for the transient evolution of 
the DSTF model from a random initial configuration to steady-state,
focusing on the trajectories of individual cars.
Using the framework of first-passage processes  
we calculated the distribution 
$P(T_{\rm FS}=t)$
of FS times,
which is the probability that a randomly selected car will be stopped for the 
first time at time $t$.
We also calculated the probability $P_{\rm S}(t)$ that a randomly selected car
will be stopped at time $t$.
For $0 < p < 1/2$, where the system evolves into an FFP state,
we also calculated the distribution 
$P(T_{\rm LS}=t)$
of LS times,
which is the probability that a randomly selected car will stop for the
last time at time $t$.
In this regime, we analyzed the relation between the LS time and the 
number of stopping events $N_{\rm S}$ 
which take place up to that time.
We presented closed-form expressions 
for the joint distribution $P(T_{\rm LS}=t,N_{\rm S}=n)$,
for the two conditional distributions that emanate from it
and for the marginal distribution $P(N_{\rm S}=n)$.
These results provide insight on the time scales of congestion
and relaxation in deterministic traffic flow.
In a broader context, they provide insight on complex relaxation processes
that involve many interacting particles, 
such as the processes that appear in deterministic surface growth
\cite{Krug1988,Katzav2004}.
 
It would be interesting to consider a broader class of initial conditions,
which may include correlations between the locations of different cars.
It would also be interesting to explore the effect of finite system sizes
and different boundary conditions on the statistical properties of
car trajectories. 
Another natural direction for future work is to extend the analysis presented in this paper
to the deterministic multi-speed family of 1D
CA models of traffic flow, referred to as the Fukui-Ishibashi (FI) model
\cite{Fukui1996}. 
The FI model is a deterministic CA model with 
synchronous update, in which each car advances by $v=\min \{ d,v_{\rm max} \}$,
where $d$ is the number of vacant cells ahead of the car and $v_{\rm max}$
is the maximum velocity.
The FI model provides a more realistic description of 1D traffic flow
in the deterministic limit.

\appendix

\section{Derivation of a useful identity}

Consider the quantity

\begin{equation}
Q_t(p) = \binom{2t}{t-1} \frac{  \left[ p(1-p) \right]^{t+1}  }{t}
\ _2 F_1 \left( \left.
\begin{array}{c}
2, 2t+1 \\
t+2
\end{array}
\right| p 
\right).
\label{eq:C_definition}
\end{equation}
 
\noindent
In this Appendix we show that the following identity holds

\begin{equation}
Q_t(p) - Q_t(1-p) = 2p - 1.
\label{eq:C_identity}
\end{equation}

\noindent
Applying the transformation presented in equation 15.8.12 in Ref. 
\cite{Olver2010} to the hypergeometric function on the right-hand side
of Eq. (\ref{eq:C_definition}), we obtain

\begin{equation}
_2F_1 \left( \left.
\begin{array}{c}
2, 2t+1 \\
t+2
\end{array}
\right| p 
\right) 
=
\left( \frac{1}{1-p} \right)^{ t+1 }
\ _2 F_1 \left( \left.
\begin{array}{c}
1-t, t \\
t+2
\end{array}
\right| p 
\right).
\label{eq:HG}
\end{equation}

\noindent
Note that the first upper index of the hypergeometric function on 
the right-hand side is zero for $t=1$ and negative for $t \ge 2$.
This implies that this hypergeometric function represents a (finite)
polynomial in $p$.
In fact, using equations 15.9.1 and 18.5.7 in Ref. \cite{Olver2010},
this hypergeometric function can be expressed 
in the form

\begin{equation}
_2F_1 \left( \left.
\begin{array}{c}
1-t, t \\
t+2
\end{array}
\right| p 
\right)
=
\frac{  P_{t-1}^{ (t+1,-t-1) } (1-2p)  }{ \binom{2t}{t-1} },
\label{eq:2F1J}
\end{equation}

\noindent
where $P_n^{ (\alpha,\beta) }(x)$ is the Jacobi polynomial
\cite{Olver2010}.
Inserting the right-hand side of Eq. (\ref{eq:2F1J})
into Eq. (\ref{eq:HG}), we obtain

\begin{equation}
_2F_1 \left( \left.
\begin{array}{c}
2, 2t+1 \\
t+2
\end{array}
\right| p 
\right) 
=
\frac{(1-p)^{ - t - 1 } }{ \binom{2t}{t-1} } P_{t-1}^{ (t+1,-t-1) } (1-2p).
\label{eq:HG2}
\end{equation}

\noindent
Inserting the right-hand side of Eq. (\ref{eq:HG2})
into Eq. (\ref{eq:C_definition}),
we obtain

\begin{equation}
Q_t(p) = \frac{p^{t+1}}{t} P_{t-1}^{(t+1,-t-1)} (1-2p).
\label{eq:Qtp}
\end{equation}

\noindent
Below we express the Jacobi polynomial as an explicit finite sum,
in order to highlight its symmetry around $p=1/2$.
To this end we use the identity
[equation 18.5.8 in Ref. \cite{Olver2010}]

\begin{equation}
P_{n}^{(\alpha,\beta)}(x) = \sum_{s=0}^{n} \binom{n+\alpha}{n-s} \binom{n+\beta}{s}
\left( \frac{x-1}{2} \right)^s \left( \frac{x+1}{2} \right)^{n-s}.
\label{eq:Jsum}
\end{equation}

\noindent
Inserting $n=t-1$, $\alpha=t+1$, $\beta=-t-1$ and $x=1-2p$
into Eq. (\ref{eq:Jsum}), we obtain

\begin{equation}
P_{t-1}^{ (t+1,-t-1) } (1-2p) =
\sum_{s=0}^{t-1} \binom{2t}{t-1-s} \binom{-2}{s} (-p)^s (1-p)^{t-1-s}.
\end{equation}

\noindent
From identity 17.2.30 in Ref. 
\cite{Olver2010}, 
with $q=1$,
we get

\begin{equation}
\binom{-2}{s} (-1)^s = s+1,
\end{equation}

\noindent
which is valid for all integers $s \ge 0$.
Using this result,
we obtain

\begin{equation}
P_{t-1}^{ (t+1,-t-1) } (1-2p) =
\sum_{s=0}^{t-1} \binom{2t}{t-1-s}  (s+1) p^s  (1-p)^{t-1-s}.
\end{equation}

\noindent
Replacing the summation index $s$ by $t-1-s$,
we obtain

\begin{equation}
P_{t-1}^{ (t+1,-t-1) } (1-2p) =
\sum_{s=0}^{t-1} \binom{2t}{s}  (t-s) p^{t-1-s}  (1-p)^{s}.
\end{equation}

\noindent
Using the identity

\begin{equation}
(t-s) \binom{2t}{s} =
t \binom{2t-1}{s} - t \binom{2t-1}{s-1},
\end{equation}

\noindent
we obtain

\begin{eqnarray}
P_{t-1}^{ (t+1,-t-1) } (1-2p) &=&
t \sum_{s=0}^{t-1} \binom{2t-1}{s}  p^{t-1-s}   (1-p)^{s} 
\nonumber \\
&-&
t \sum_{s=0}^{t-1} \binom{2t-1}{s-1}  p^{t-1-s}  (1-p)^{s}.
\label{eq:PJ1}
\end{eqnarray}

\noindent
Shifting the summation index in the second sum on the 
right-hand side of Eq. (\ref{eq:PJ1}) from $s$ to $s-1$, we obtain

\begin{eqnarray}
P_{t-1}^{ (t+1,-t-1) } (1-2p) &=&
t \sum_{s=0}^{t-1} \binom{2t-1}{s}  p^{t-1-s}   (1-p)^{s} 
\nonumber \\
&-&
t \sum_{s=0}^{t-2} \binom{2t-1}{s}  p^{t-2-s}  (1-p)^{s+1}.
\label{eq:PJ2}
\end{eqnarray}

\noindent
Inserting the right-hand side of Eq. (\ref{eq:PJ2}) into Eq. (\ref{eq:Qtp})
and rearranging terms,
we obtain
 
\begin{eqnarray}
Q_t(p) &=& 
\binom{2t-1}{t-1} p^{t+1} (1-p)^{t-1}
\nonumber \\
&+&
(2p-1) \sum_{s=0}^{t-2} \binom{2t-1}{s} p^{2t-1-s} (1-p)^s.
\label{eq:Ctp1}
\end{eqnarray} 

\noindent
Replacing $p$ in Eq. (\ref{eq:Ctp1}) by $1-p$, we obtain

\begin{eqnarray}
Q_t(1-p) &=& 
\binom{2t-1}{t-1} (1-p)^{t+1}  p^{t-1}
\nonumber \\
&+&
(1-2p) \sum_{s=0}^{t-2} \binom{2t-1}{s} (1-p)^{2t-1-s}  p^s.
\label{eq:Ctp2}
\end{eqnarray}

\noindent
Subtracting Eq. (\ref{eq:Ctp2}) from Eq. (\ref{eq:Ctp1}), we obtain

\begin{eqnarray}
Q_t(p) -  Q_t(1-p) &=&   
\left[ p^2 - (1-p)^2 \right]
\binom{2t-1}{t-1} p^{t-1} (1-p)^{t-1}
\nonumber \\
&+& 
(2p-1)  
 \sum_{s=0}^{t-2} \binom{2t-1}{s} p^{2t-1-s} (1-p)^s
 \nonumber \\
 &+&
 (2p-1)
\sum_{s=0}^{t-2} \binom{2t-1}{s} (1-p)^{2t-1-s}  p^s.
\label{eq:Ctp3}
\end{eqnarray} 

\noindent
Reindexing the second sum on the right-hand side 
and rearranging terms ($s \rightarrow 2t-s-1$),
we obtain

\begin{eqnarray}
Q_t(p) - Q_t(1-p) &=&   
(2p-1)
\binom{2t-1}{t-1} p^{t-1} (1-p)^{t-1}
\nonumber \\
&+& 
(2p-1)  
\sum_{s=0}^{t-2} \binom{2t-1}{s} p^{2t-1-s} (1-p)^s
\nonumber \\
&+&
(2p-1)
\sum_{s=t+1}^{2t-1} \binom{2t-1}{2t-1-s} (1-p)^{s}  p^{2t-1-s}.
\label{eq:Ctp4}
\end{eqnarray} 

\noindent
The two sums on the right-hand side add up to almost the full binomial expansion
of $[p + (1-p)]^{2t-1} = 1$, apart from two terms ($s=t-1$ and $s=t$).
More specifically

\begin{eqnarray}
&& \sum_{s=0}^{t-2} \binom{2t-1}{s} p^{2t-1-s} (1-p)^s
+
\sum_{s=t+1}^{2t-1} \binom{2t-1}{2t-1-s} (1-p)^{s}  p^{2t-1-s}
\nonumber \\
&& = 1 - \binom{2t-1}{t-1} p^{t} (1-p)^{t-1} - \binom{2t-1}{t} p^{t-1} (1-p)^t.
\label{eq:binoms}
\end{eqnarray}

\noindent
Inserting the right-hand side of Eq. (\ref{eq:binoms}) into Eq. (\ref{eq:Ctp4})
and rearranging terms, we obtain

\begin{equation}
Q_t(p) - Q_t(1-p) = 2 p - 1.
\end{equation}

\section{The generating function of the distribution of last-stopping times}

In this Appendix we carry out the calculation of the generating function

\begin{equation}
R(z) = \sum_{t=1}^{\infty} z^t P(T_{\rm LS}=t).
\label{eq:Rz1}
\end{equation}

\noindent
To this end, we express $P(T_{\rm LS}=t)$ in terms of the
Lobb numbers and insert it into Eq. (\ref{eq:Rz1}),
which yields

\begin{equation}
R(z) = \frac{1-2p}{p} \sum_{t=1}^{\infty}
(zp)^t \sum_{\ell=0}^{t-1} L_{\frac{t-1+\ell}{2},\frac{t-1-\ell}{2}}
(1-p)^{\ell}.
\label{eq:Rz0}
\end{equation}

\noindent
In order to proceed, we use some results for the 
generalized Fuss-Catalan numbers
(also called Raney numbers), defined by
\cite{Chou2018}

\begin{equation}
F_r(n,k) = \frac{k}{rn + k} \binom{rn+k}{n}.
\label{eq:Fmnk}
\end{equation}

\noindent
The generating function of the generalized Fuss-Catalan numbers
is given by
\cite{Chou2018}

\begin{equation}
\sum_{n=0}^{\infty} F_r(n,k) z^n = \left[ F_r(z) \right]^k,
\label{eq:sumFmnk}
\end{equation}

\noindent
where $F_r(z)$ is the generating function of the case $k=1$,
namely

\begin{equation}
F_r(z) = \sum_{n=0}^{\infty} F_r(n,1) z^n.
\end{equation}

\noindent
In the following we will be interested in the case of $r=2$.
Inserting $r=2$ and $k=1$ into Eq. (\ref{eq:Fmnk}),
it is found that 

\begin{equation}
F_2(n,1) = C_n,
\end{equation}

\noindent
where $C_n$ is the Catalan number.
This implies that $F_2(z)$ is in fact equal to the generating 
function of the Catalan numbers, denoted by $C(z)$,
namely
\cite{Chou2018}

\begin{equation}
F_2(z) = C(z) = \frac{ 1 - \sqrt{ 1 - 4z } }{2z}.
\label{eq:F2z}
\end{equation}

\noindent
Replacing $F_2(z)$ by $C(z)$ on the right-hand side of Eq. (\ref{eq:sumFmnk})
with $r=2$,
we obtain

\begin{equation}
\sum_{n=0}^{\infty} F_2(n,k) z^n = \left[ C(z) \right]^k.
\label{eq:sumF2nk}
\end{equation}

\noindent
In fact, for $r=2$ the generalized Fuss-Catalan numbers are
reduced to the Lobb numbers, namely

\begin{equation}
L_{n,m} = F_2(n-m,2m+1).
\label{eq:LF2}
\end{equation}

\noindent
Replacing the Lobb number on the right-hand side of Eq. (\ref{eq:Rz0})
by the corresponding generalized Fuss-Catalan number, we obtain

\begin{equation}
R(z) = \frac{1-2p}{p} \sum_{t=1}^{\infty}
(zp)^t \sum_{\ell=0}^{t-1} 
(1-p)^{\ell}
F_2(\ell,t-\ell).
\label{eq:Rz2}
\end{equation}

\noindent
Exchanging the order of summations, we obtain

\begin{equation}
R(z) = \frac{1-2p}{p} 
\sum_{\ell=0}^{\infty} 
\sum_{t=\ell+1}^{\infty}
(zp)^t 
(1-p)^{\ell}
F_2(\ell,t-\ell).
\label{eq:Rz3}
\end{equation}

\noindent
Shifting the summation index $t$ by $\ell$,
we obtain

\begin{equation}
R(z) = \frac{1-2p}{p} 
\sum_{\ell=0}^{\infty} 
\sum_{t=1}^{\infty}
(zp)^{t+\ell} 
(1-p)^{\ell} 
F_2(\ell,t).
\label{eq:Rz4}
\end{equation}

\noindent
Exchanging the order of summation again,
we obtain

\begin{equation}
R(z) = \frac{1-2p}{p} 
\sum_{t=1}^{\infty}
(zp)^{t} 
\sum_{\ell=0}^{\infty} 
[zp(1-p)]^{\ell}
F_2(\ell,t).
\label{eq:Rz5}
\end{equation}

\noindent
Expressing the sum over $\ell$ in terms of the
generating function shown in Eq. (\ref{eq:sumF2nk}),
we obtain

\begin{equation}
R(z) = \frac{1-2p}{p} 
\sum_{t=1}^{\infty}
(zp)^{t} 
\left\{ C \left[ z p (1-p) \right] \right\}^{t}.
\label{eq:Rz6}
\end{equation}

\noindent
Summing this geometric series yields

\begin{equation}
R(z) = \frac{1-2p}{p}  \frac{ z p C \left[ z p (1-p) \right] }{1 - z p C \left[ z p (1-p) \right] }.
\label{eq:Rz7}
\end{equation}

\noindent
Inserting $C(z)$ from Eq. (\ref{eq:F2z}) into Eq. (\ref{eq:Rz7}),
we obtain

\begin{equation}
R(z) = \frac{1-2p}{p} \frac{ 1 - \sqrt{1-4zp(1-p)} }{ 1-2p+\sqrt{1-4zp(1-p)} }.
\label{eq:Rz8}
\end{equation}


\clearpage
\newpage


\begin{thebibliography}{10}



\bibitem{Wolfram1983}
S. Wolfram,
Statistical mechanics of cellular automata,
{\it Rev. Mod. Phys.} {\bf 55}, 601 (1983).

\bibitem{Wolfram1984}
S. Wolfram,
Universality and complexity in cellular automata,
{\it Physica D} {\bf 10}, 1 (1984).



\bibitem{Biham1992}
O. Biham, A.A. Middleton and D. Levine,
Self organization and a dynamical transition in traffic flow models,
{\it Phys. Rev. A} {\bf 46}, R6124 (1992).



\bibitem{Maerivoet2005}
S. Maerivoet and B. De Moor,
Cellular automata models of road traffic,
{\it Physics Reports} {\bf 419}, 1 (2005).




\bibitem{Chowdhury2000}
D. Chowdhury, L. Santen and A. Schadschneider,
Statistical physics of vehicular traffic and some related systems,
{\it Physics Reports} {\bf 329}, 199 (2000).

\bibitem{Helbing2001}
D. Helbing,
Traffic and related self-driven many-particle systems,
{\it Rev. Mod. Phys.} {\bf 73}, 1067 (2001).

\bibitem{Schadschneider2002}
A. Schadschneider,
Traffic flow: a statistical physics point of view,
{\it Physica A} {\bf 313}, 153 (2002).

\bibitem{Nagatani2002}
T. Nagatani,
The physics of traffic jams,
{\it Rep. Prog. Phys.} {\bf 65}, 1331 (2002).



\bibitem{Nagel1992}
K. Nagel and M. Schreckenberg,
A cellular automaton model for freeway traffic,
{\it J. de Physique I} {\bf 2}, 2221 (1992).

\bibitem{Nagel1993}
K. Nagel and H. J. Herrmann,
Deterministic models for traffic jams,
{\it Physica A} {\bf 199}, 254 (1993).

\bibitem{Schadschneider1993}
A. Schadschneider and M. Schreckenberg,
Cellular automaton models and traffic flow,
{\it J. Phys. A} {\bf 26}, L679 (1993).

\bibitem{Schreckenberg1995}
M. Schreckenberg, A. Schadschneider, K. Nagel and N. Ito,
Discrete stochastic models for traffic flow,
{\it Phys. Rev. E} {\bf 51}, 2939 (1995).



\bibitem{Kerner1996}
B.S. Kerner and H. Rehborn,
Experimental properties of complexity in traffic flow,
{\it Phys. Rev. E} {\bf 53}, 4275 (1996).


\bibitem{Kerner2002}
B.S. Kerner, S.L. Klenov and D.E. Wolf,
Cellular automata approach to three-phase traffic theory,
{\it J. Phys. A} {\bf 35}, 9971 (2002).



\bibitem{Krug1988}
J. Krug and H. Spohn,
Universality classes for deterministic surface growth,
{\it Phys. Rev. A} {\bf 38}, 4271 (1988). 

\bibitem{Sasvari1997}
M. Sasv\'ari, and J. Kert\'esz,
Cellular automata models of single-lane traffic,
{\it Phys. Rev. E} {\bf 56}, 4104 (1997).

\bibitem{Belitsky2005}
V. Belitsky and P. A. Ferrari,
Invariant measures and convergence properties
for cellular automaton 184 and related processes,
{\it J. Stat. Phys.} {\bf 118}, 589 (2005).

\bibitem{Boccara1998}
N. Boccara and H. Fuk\'s,
Cellular automaton rules conserving the number of active sites,
{\it J. Phys. A} {\bf 31}, 6007 (1998).

\bibitem{Boccara2002}
N. Boccara and H. Fuk\'s, 
Number-conserving cellular automaton rules, 
{\it Fundam. Inf.} {\bf 52}, 1 (2002).

\bibitem{Fuks1997}
H. Fuk\'s,
Solution of the density classification problem with two cellular automata rules,
{\it Phys. Rev. E} {\bf 55}, R2081 (1997).

\bibitem{Fuks1999}
H. Fuk\'s,
Exact results for deterministic cellular automata traffic models,
{\it Phys. Rev. E} {\bf 60}, 197 (1999).



\bibitem{Fuks2023}
H. Fuk\'s,
{\it Solvable Cellular Automata: Methods and Applications}
(Springer, Cham, 2023).

\bibitem{Jha2025}
A. Jha, K. Wiesenfeld, G. Lee and J. Laval,
Simple traffic model as a space-time clustering phenomenon,
{\it Phys. Rev. E} {\bf 112}, 054104 (2025).




\bibitem{Redner2001}
S. Redner,
{\it A Guide to First Passage Processes}
(Cambridge University Press, Cambridge, 2001).




\bibitem{Spitzer1970}
F. Spitzer,
Interaction of Markov processes,
{\it Advances in Mathematics} {\bf 5}, 246 (1970).


\bibitem{Derrida1992}
B. Derrida, E. Domany and D. Mukamel, 
An exact solution of a one-dimensional
asymmetric exclusion model with open boundaries,
{\it J. Stat. Phys.} {\bf 69}, 667  (1992).


\bibitem{Derrida1998}
B. Derrida,
An exactly soluble non-equilibrium system:
The asymmetric simple exclusion process,
{\it Phys. Rep.} {\bf 301}, 65 (1998).  


\bibitem{Kardar1986}
M. Kardar, G. Parisi and Y.-C. Zhang, 
Dynamic Scaling of Growing Interfaces, 
{\it Phys. Rev. Lett.} {\bf 56}, 889 (1986). 

\bibitem{Schutz1993}
G.M. Sch\"utz,  
Time-dependent correlation functions in a one-dimensional 
asymmetric exclusion process, 
{\it J. Stat. Phys.} {\bf 71}, 471 (1993).

\bibitem{Rajewsky1998}
N. Rajewsky, L. Santen, A. Schadschneider and M. Schreckenberg, 
The asymmetric exclusion process: comparison of update procedures, 
{\it J. Stat. Phys.} {\bf 92}, 151 (1998).



\bibitem{Belitsky2011a}
V. Belitsky and G.M. Sch\"utz,
Cellular automaton model for molecular traffic jams,
{\it J. Stat. Mech.} P07007 (2011).

\bibitem{Belitsky2011b}
V. Belitsky and G.M. Sch\"utz,
Microscopic position and structure of a shock in CA 184,
{\it J. Phys. A} {\bf 44}, 445003 (2011).



\bibitem{Mounaix2020}
P. Mounaix, S. N. Majumdar and G. Schehr,
Statistics of the number of records for random walks and L\'evy flights on a 1D lattice,
{\it J. Phys. A} {\bf 53}, 415003 (2020).



\bibitem{Majumdar2024}
S.N. Majumdar and G. Schehr,
{\it Statistics of Extremes and Records in Random Sequences}
(Oxford University Press, Oxford, 2024).


\bibitem{Flajolet2009}
P. Flajolet and R. Sedgewick,
{\it Analytic Combinatorics}
(Cambridge University Press, Cambridge, 2009).



\bibitem{Audibert2010}
P. Audibert,
{\it Mathematics for Informatics and Computer Science}
(ISTE and Hoboken, London, 2010).



\bibitem{Stanley2015}
R.P. Stanley,
{\it Catalan Numbers}
(Cambridge University Press, Cambridge, 2015).


\bibitem{Koshy2009}
T. Koshy,
{\it Catalan Numbers with Applications}
(Oxford University Press, Oxford, 2009).






\bibitem{Deutsch1999}
E. Deutsch,
Dyck path enumeration,
{\it Discrete Mathematics} {\bf 204}, 167 (1999).
 





\bibitem{Gruda2025}
M. Gruda, O. Biham, E. Katzav and R. K\"uhn,
The joint distribution of first return times and of 
the number of distinct sites visited by a 1D random walk 
before returning to the origin,
{\it J. Stat. Mech.} 013203 (2025).
 


\bibitem{Klinger2022}
J. Klinger, A. Barbier-Chebbah, R. Voituriez and O. B\'enichou,
Joint statistics of space and time exploration of one-dimensional random walks,
{\it Phys. Rev. E} {\bf 105}, 034116 (2022).



\bibitem{Polya1921}
G. P\'olya,
\"{U}ber eine aufgabe der wahrscheinlichkeitsrechnung betreffend die irrfahrt im strassennetz
{\it Mathematische Annalen} {\bf 84}, 149 (1921).



\bibitem{Olver2010}
F.W.J. Olver, D.M. Lozier, R.R. Boisvert and C.W. Clark,
{\it NIST Handbook of Mathematical Functions}
(Cambridge University Press, Cambridge, 2010).

\bibitem{Feller1971a}
W. Feller,
{\it An Introduction to Probability Theory and its Applications: Vol. I}
(Wiley, New York, 1971). 

\bibitem{Feller1971b}
W. Feller,
{\it An Introduction to Probability Theory and its Applications: Vol. II}
(Wiley, New York, 1971).  




\bibitem{Majumdar2012}
S.N. Majumdar, G. Schehr and G. Wergen,
Record statistics and persistence for a random walk with a drift,
{\it J. Phys. A} {\bf 45}, 355002 (2012).



\bibitem{Lobb1999}
A. Lobb,
Deriving the $n$th Catalan number,
{\it The Mathematical Gazette} {\bf 83}, 109 (1999).



\bibitem{Koshy2012}
T. Koshy,
Lobb's generalisation of Catalan's parenthesisation problem revisited,
{\it The Mathematical Gazette}  {\bf 96}, 56 (2012).











\bibitem{Toussaint1983}
D. Toussaint and F. Wilczek, 
Particle-antiparticle annihilation in diffusive motion,
{\it J. Chem. Phys.} {\bf 78}, 2642 (1983).

\bibitem{Kang1984}
K. Kang and S. Redner,
Scaling approach for the kinetics of recombination processes,
{\it Phys. Rev. Lett.} {\bf 52}, 955 (1984).

\bibitem{Bramson1988}
M. Bramson and J.L. Lebowitz,
Asymptotic behavior of densities in diffusion-dominated annihilation reactions,
{\it Phys. Rev. Lett.} {\bf 61}, 2397 (1988);
Erratum: {\it Phys. Rev. Lett.} {\bf 62}, 694 (1989).
 


\bibitem{Krapivsky2010}
P. Krapivsky, E. Ben-Naim and S. Redner, 
{\it A Kinetic View of Statistical Physics}
(Cambridge University Press, Cambridge, 2010).


\bibitem{Elskens1985}
Y. Elskens and H.L. Frisch,
Annihilation kinetics in the one-dimensional ideal gas,
{\it Phys. Rev. A} {\bf 31}, 3812 (1985).



\bibitem{Paris2013}
R.B. Paris, 
Asymptotics of the gauss hypergeometric function with large parameters I, 
{\it Journal of Classical Analysis} {\bf 2}, 183 (2013).
 


\bibitem{Hertzberg2026}
G. Hertzberg Rabinovich, O. Biham and E. Katzav,
Structure and dynamics in the low-density phase of a two-dimensional 
cellular automaton model of traffic flow,
{\it Phys. Rev. E} {\bf 113}, 014127 (2026).
 
 


\bibitem{Katzav2004}
E. Katzav and M. Schwartz,
What is the connection between ballistic deposition and the 
Kardar-Parisi-Zhang equation?,
{\it Phys. Rev. E} {\bf 70}, 061608 (2004).
 
 


 
 
 

\bibitem{Fukui1996}
M. Fukui and Y. Ishibashi, 
Traffic flow in 1D cellular automaton model including cars
moving with high speed,
{\it J. Phys. Soc. Jpn.} {\bf 65}, 1868 (1996).



\bibitem{Ross2020}
S.M. Ross, 
{\it A First Course in Probability, 10th Edition}
(Pearson, Upper Saddle River, NJ, 2020), 
page 399.



\bibitem{Chou2018}
W.-S. Chou, T.-X. He and P. J.-S. Shiue,
On the primality of the generalized Fuss-Catalan numbers,
{\it Journal of Integer Sequences} {\bf 21}, 18.2.1 (2018).




\end{thebibliography}
\end{document}